\documentclass[preprint2]{aastex}

\slugcomment{}

\shorttitle{H$_2$ in Intermediate-Velocity Clouds}
\shortauthors{Richter et al.}

\begin{document}

\title{A FUSE Survey of Molecular Hydrogen in Intermediate-Velocity Clouds
in the Milky Way Halo}


\author{Philipp Richter\altaffilmark{1,2}, Bart P. Wakker\altaffilmark{1}, 
Blair D. Savage\altaffilmark{1}}
\and
\author{Kenneth R. Sembach\altaffilmark{3}} 

\altaffiltext{1}{Department of Astronomy University of Wisconsin-Madison,
475 N. Charter Street, Madison, WI\,53706; richter@astro.wisc.edu}
\altaffiltext{2}{Osservatorio Astrofisico di Arcetri, Largo E. Fermi 5, 50125 Florence, Italy}
\altaffiltext{3}{Space Telescope Science Institute, 3700 San Martin Drive, Baltimore MD\,21218}



\begin{abstract}

{\it Far Ultraviolet Spectroscopic Explorer} 
(FUSE) data is used to investigate
the molecular hydrogen (H$_2$) content of intermediate-velocity
clouds (IVCs) in the lower halo of the Milky Way. 
We analyze interstellar absorption towards 56 
(mostly extragalactic) background sources
to study H$_2$ absorption in the Lyman- and Werner bands in 
61 IVC components at H\,{\sc i} column densities 
$\geq 10^{19}$ cm$^{-2}$.
For data with good S/N ($\sim 9$ per resolution element and higher),
H$_2$ in IVC gas is convincingly detected in 14 cases at column densities
varying between $\sim 10^{14}$ and $\sim 10^{17}$ cm$^{-2}$. 
We find an additional 17 possible H$_2$ detections in IVCs in FUSE 
spectra with lower S/N. The molecular hydrogen fractions, $f$, 
vary between $10^{-6}$ and $10^{-3}$, implying a dense, mostly neutral
gas phase that is probably related to the Cold Neutral Medium (CNM) in 
these clouds. If the H$_2$ stays in formation-dissociation equlibrium,
the CNM in these clouds can be characterized 
by compact ($D\sim0.1$ pc) filaments with
volume densities on the order of   
$n_{\rm H} \sim30$ cm$^{-3}$. The relatively high detection
rate of H$_2$ in IVC gas implies that the CNM in these
clouds is ubiquitous. 
More dense regions
with much higher molecular fractions 
may exist, but it would be difficult to detect them in absorption
because of their small size. 
\end{abstract}


\section{Introduction}

Direct observational information 
on the most abundant interstellar 
molecule in the Universe, molecular hydrogen 
(H$_2$), is difficult to 
obtain. 
In diffuse interstellar gas, H$_2$ can be
studied directly by way of far-ultraviolet (FUV) absorption
spectroscopy in its electronic
Lyman and Werner bands in the wavelength
regime between $912$ and $1130$ \AA\,
toward UV-bright background sources,
such as early type stars, supernovae,
quasars (QSOs) and active galactic nuclei (AGNs).
In the low-redshift Universe, this
bandpass is accessible only with
space-based instrumentation.
Previous FUV satellites, such as
{\it Copernicus} and ORFEUS, observed
only very nearby stars and a few 
selected distant objects.
An overall assessment of the H$_2$ 
content of the diffuse ISM outside
the Milky Way disk and the Magellanic
Clouds was not possible before
1999, when the
{\it Far Ultraviolet Spectroscopic Explorer}
(FUSE) was launched. The high sensitivity of
FUSE allows observations  of
H$_2$ absorption in
regions outside the Milky Way disk
towards a large number of extragalactic
background sources. Such observations
reveal the abundance and 
properties of diffuse molecular gas
in a variety of different environments.

Particularly promising regions
for studies of H$_2$ are the Milky Way halo and
its neutral gas clouds that move with
intermediate ($30 \leq |v_{\rm LSR}| \leq 90$ km\,s$^{-1}$; IVCs)
and high ($|v_{\rm LSR}| \geq 90$ km\,s$^{-1}$; HVCs) radial
velocities several kiloparsecs above the
Galactic plane. Recent studies with
FUSE and the {\it Space Telescope Imaging
Spectrograph} (STIS) have shown that
the chemical composition of these
clouds is not uniform, with metallicities varying
between $\sim 0.1$ and $\sim 1.0$ solar
(Wakker et al.\,1999; Richter
et al.\,2001a, 2001b; Lu et al.\,1998;
Gibson et al.\,2000, 2001; Wakker 2001). The results
imply a separation into three different
groups of Galactic halo clouds:
(1) Gas that has been ejected out of
the Milky Way disk by a ``Galactic
Fountain'' (Shapiro \& Field 1976) or
similar processes, having $\sim 1.0$ solar
abundances. All of the IVCs probably fall into
this category. (2) Gas that has been
brought into the Milky Way halo due to
tidal interactions with the Magellanic Clouds,
having metallicities similar to that of
the Magellanic Clouds ($\sim 0.3$ solar for the
Magellanic Stream).
(3) Low-metallcity gas ($\sim 0.1$ solar
abundances) that is infalling into
the Galactic halo from intergalactic
space (e.g., HVC Complex C).
Because of the varying metal and dust content
of IVCs and HVCs, and the
reduced UV radiation field above 
the Galactic plane,
IVCs and HVCs may serve as important 
interstellar laboratories to study H$_2$
formation- and dissociation processes under
conditions that are also found in more
distant regions of the Universe (e.g., in
intergalactic Ly\,{$\alpha$} absorbers) 
for which detailed H$_2$
absorption spectroscopy
is often impossible.   

Except for some IVCs that are close
to the disk (the Draco cloud - 
Mebold et al.\,1985, and IV\,21 - Weiss et al.\,1999,
Heithausen et al.\,2001), no CO emission
or absorption has been detected in
Galactic halo clouds so far.
H$_2$ was found for the
first time in HVC gas (Richter
et al.\,1999) and IVC gas (Gringel et al.\,2000)
in data of the ORFEUS satellite,
followed by further detections with
FUSE (Sembach et al.\,2001a; Richter et al.\,2001a).
In a previous paper (Richter et al.\,2001c), we
studied H$_2$ absorption in two prominent
metal-poor HVCs,
the Magellanic Stream and Complex C.
While H$_2$ is clearly detected in the
Magellanic Stream (see also Sembach et al.\,2001a), 
there is no evidence
for the presence of H$_2$ in the metal- and
dust-poor Complex C (Richter et al.\,2001b).
This implies that the formation and maintainance of 
diffuse H$_2$ in halo clouds is tightly 
correlated to the availability of interstellar
dust grains, on whose surface the H$_2$ formation
proceeds most efficiently (see, e.g., Shull \& Beckwith 1982).

In this paper we investigate the distribution
of H$_2$ in IVCs toward 56 background sources.
The data sample nearly all of the most
prominent, large IVCs located predominantly in
the northern Galactic sky. These IVCs
are: the Intermediate-Velocity Arch (IV Arch),
the Low-Latitude Intermediate-Velocity Arch
(LLIV Arch), and the Intermediate-Velocity Spur
(IV Spur) in the northern Galactic sky, and 
Complex gp and the Pegasus-Pisces Arch (PP Arch) 
in the southern sky.
As summarized by Wakker (2001), all of these
IVCs are known to have roughly solar, or slightly
sub-solar 
metallicities, and are located
between $0.3$ and $2.1$ kpc away from the
Galactic plane; thus they are in the ``lower'' Galactic halo.
Distance brackets for IVCs are derived by comparing 
IVC absorption line detections and non-detections 
toward stars at known distances and $z$-heights
(see Wakker 2001 for details).  

In Figure 1 we show the negative velocity H\,{\sc i} IVC sky 
based on 21cm data from the Leiden-Dwingeloo
Survey (LDS; Hartmann \& Burton 1997) 
at H\,{\sc i} column densities 
$N($H\,{\sc i}$)\geq 2 \times 10^{19}$ cm$^{-2}$.
The various IVC complexes are identified and
explained in detail by Wakker (2001). 
Intermediate-velocity H\,{\sc i} above this column density
covers a significant portion of the sky ($\sim 35$ 
percent for $30 \leq |v_{\rm LSR}| \leq 90$ km\,s$^{-1}$), 
but it is very likely
that the sky-coverage is much higher for low-column density
H\,{\sc i} gas below the detection limit of the 21cm observations.
The high sky coverage of IVC gas in the halo or disk-halo
interface underlines the importance of the IVC phenomenon
to an understanding of the ISM of the Milky Way, although
there do not exist many systematic studies of IVCs in the
literature. As pointed
out by Kulkarni \& Fich (1985), IVCs actually contain much of
the kinetic energy of the total Galactic ISM. 
Moreover, IVCs appear to have 
a mass circulation rate much higher than the HVCs, and
high enough to produce the O\,{\sc vi} seen in the thick
Galactic disk, if the IVCs represent the return flow
of a Galactic Fountain (see Savage et al.\,2002).

Some of the IVCs have been studied in detail
using FUSE absorption line data (LLIV Arch, Richter
et al.\,2001a; IV Arch, Richter et al.\,2001b).
The elemental depletion pattern observed in IVCs 
implies that these
clouds contain dust grains, but the surface
properties of these grains, important for
the H$_2$ formation, are unknown.
With respect to their velocities, abundances, and distances,
it appears likely that IVCs represent the
cooled return flow of a Galactic Fountain
(Shapiro \& Field 1976).
One of our sight lines (Mrk\,876) samples
the Draco cloud ($z=0.3-0.8$ kpc),
one of the two
IVCs at lower $z$ that is known to contain
CO. It is not yet
clear whether the Draco cloud falls into the same
class of objects as the other IVCs
cited above (possibly, as a final stage
of the Galactic Fountain process), or
whether it represents a Galactic molecular
cloud at unusually high Galactic latitudes
and $z$-height.
Our FUSE data also sample a variety
of other small IVCs in the northern and southern
Galactic sky. Distances and metallicities
for these IVCs are unknown, but we have 
formally included these clouds in our 
sample, based on criteria that are described in 
\S4.

This paper is organized as follows:
In \S2 we discuss
the FUSE observations and the data handling.
In \S3 we analyze in detail molecular 
hydrogen absorption in the IV Arch in the FUSE spectrum of 
the quasar PG\,1351+640. In
\S4 we present a survey of 
H$_2$ in IVCs toward
56 (mostly extragalactic) background sources.
A comparison
between the H$_2$ and the H\,{\sc i}
sky at intermediate velocities is given
in \S5.
In \S6 and \S7 we investigate the origin 
and physical conditions of the diffuse
molecular gas in the lower halo.
A discussion of our observations
is presented in \S8. We summarize our study
in \S9.

\section{Observations and Data Handling}

In this paper we are making use of FUSE
absorption line data of $\sim 200$ sources,
predominantly extragalactic objects
(QSOs, Seyferts, Galaxies, BLLacs), as well as
two distant ($d>5$ kpc) Galactic halo stars.
This data is available from the FUSE
survey of interstellar O\,{\sc vi} in
the thick disk of the Milky Way 
and in high-velocity clouds
(Wakker et al.\,2002; Savage et al.\,2002; 
Sembach et al.\,2002).
A detailed description of the FUSE data and
the analysis method is provided by
Wakker et al.\,(2002). We briefly
summarize the observations and the
data reduction process.
The FUSE spectra were collected in the time
between November 1999 and December 2001
as part of various Principal-Investigator
(PI) projects and Guest-Investigator (GI)
projects.
\footnote{
All the FUSE data presented are publicly
available at the MAST archive at
{\tt  http://archive.stsci.edu/fuse}.}
FUSE is equipped with four co-aligned prime-focus
telescopes, Rowland-type spectrographs, and
two microchannel plate detectors,
in total covering the wavelength range from
905 to 1187 \AA. Four independent channels are
available, two having a Al+LiF coatings (for $\lambda 
\geq 1000$ \AA\,) and two 
having SiC surfaces ($\lambda 
\leq 1100$ \AA\,). There are three
entrance apertures, LWRS ($30\farcs0 \times 30\farcs0$),
MDRS ($4\farcs0 \times 20\farcs0$), and HIRS 
($1\farcs25 \times 20\farcs0$).
Details about the instrument and its 
in-orbit performance can be found
in Moos et al.\,(2000) and Sahnow et al.\,(2000).

Most of the data that is used here 
were recorded through the
large aperture (LWRS)
and in the photon-address mode,
in which the X/Y location, the
arrival time, and the pulse height
of each detection are stored in
a photon list.
All spectra
have been reduced using the {\tt CALFUSE}
standard pipeline (v1.8.7) that corrects
for detector backgrounds, spacecraft orbital
motions, and geometrical distortions (Sahnow
et al.\,2000). 
A wavelength calibration
was made for each individual spectrum by
adjusting various atomic lines (e.g., Si\,{\sc ii}
$\lambda 1020.7$, Ar\,{\sc i} $\lambda 1048.2$)
and molecular hydrogen lines to match the
distribution of H\,{\sc i} emission along
each sight line (see Wakker et al.\,2002)
on the LSR velocity scale. We estimate
an accuracy of $\sim 10-15$ km\,s$^{-1}$ for the
velocity calibration. The spectral resolution of the
FUSE data is $\sim 20$ km\,s$^{-1}$, with slight
variations ($\sim \pm 5$ km\,s$^{-1}$) within 
the data set. The fluxes of our background
sources vary between $0.1$ and $60 \times 10^{-14}$
erg\,cm$^{-2}$\,s$^{-1}$\,\AA$^{-1}$ (see
Wakker et al.\,2002). Only a fraction
of the entire data set ($\sim 50$ percent) can actually be used 
to study interstellar absorption in the halo
at a sufficient continuum flux.
The data for each sight line were rebinned in 
an appropriate manner in order to improve the
signal-to-noise ratio (S/N). For the spectra
that are actually used to study IVC H$_2$ absorption
in this study (56 spectra), bin sizes range
between 3 and 10 pixels, corresponding
to $\sim 6$ to $\sim 20$ km\,s$^{-1}$. The S/N in the
data varies between  3 and 30 (these values 
represent the S/N per resolution element; 
see also Wakker et al.\,2002).
Continua were fitted using low-order polynomials.
Most of the continua are relatively flat, reflecting
the smooth nature of the UV spectrum of QSOs and
other extragalactic background sources. In some cases,
emission that is associated with the background object 
complicates the continuum placement, but the
corresponding continuum variations arise on a scale that
is much larger than that of the interstellar absorption lines. 
For these cases, the continuum
was fitted locally in the vicinity of each line.
\footnote{Similary, this method was used for the two stellar
background sources that have more irregulary shaped continua.}
Equivalent widths were measured by fitting multi-component
Gaussian line profiles to the data. This method is particulary
helpful to reproduce a multi-component absorption in 
a spectrum with low S/N and larger bin sizes.
Equivalent width errors are based on photon statistics
and contiuum placement errors. 

\section{H$_2$ Absorption in the IV Arch toward PG\,1351+640}

In this section we present a detailed description
and discussion of H$_2$ absorption in the 
Intermediate-Velocity Arch (IV Arch) in the
FUSE spectrum of the quasar PG\,1351+640,
which represents the best example
of molecular absorption in intermediate-velocity
halo gas. 
For most of the other sight lines presented later,
such a detailed study is not possible due to
the lower data quality and the less
pronounced nature of the H$_2$ absorption.
This section, however, may serve as an introduction
to the general analysis method that we will also use for the
other sight lines discussed in \S4. 

Figure 2 presents the sight-line structure towards
PG\,1351+640, as seen in Effelsberg H\,{\sc i} 21cm emission
data (top panel, from Wakker et al.\,2001) and in atomic
absorption from P\,{\sc ii} $\lambda 1152.8$, Si\,{\sc ii}
$\lambda 1020.7$, and Fe\,{\sc ii} $\lambda 1122.0$
(lower three panels) from FUSE data, plotted on a LSR 
radial velocity scale. The main emission and absorption
components include: (1) local Milky Way gas near
$0$ km\,s$^{-1}$, (2) the IV Arch
seen near $-50$ km\,s$^{-1}$, and (3) 
HVC Complex C at $-155$ km\,s$^{-1}$. 
Detailed maps of 21cm emission in the IV Arch and Complex C can be 
found in Kuntz \& Danly (1996) and Wakker (2001).
As the H\,{\sc i} emission profile shows, all
three major components have substructure,
for instance a weak second IVC component near
$-74$ km\,s$^{-1}$, which is also seen in atomic
absorption. The Effelsberg data yield
H\,{\sc i} column densities of $1.12 \times 10^{20}$
cm$^{-2}$ for the main IVC component at
$-50$ km\,s$^{-1}$, and $1.58 \times 10^{18}$
cm$^{-2}$ for the weaker component 
at $-74$ km\,s$^{-1}$.
The $-50$ km\,s$^{-1}$ component is 
associated with the IV\,19 core (see Kuntz \& Danly 1996),
while the $-74$ km\,s$^{-1}$ component is part
of core IV\,9.

H$_2$ absorption in the IV Arch toward PG\,1351+640
is found in 29 lines in rotational levels $J=0$ to
$3$ at a radial velocity of $v_{\rm LSR} \approx -50$ km\,s$^{-1}$,
representing the main IVC component along this sight line 
(see above)
\footnote{No H$_2$ absorption is evident in the weaker
$-74$ km\,s$^{-1}$ component; see also Table 2.}.
Figure 3 presents a portion of the FUSE spectrum
of PG\,1351+640 in the wavelength region of the 
H$_2$ Lyman $2-0$ band, together with a two-component Gaussian fit
of the data.
Figure 4 shows a selection of velocity profiles for the
rotational ground states $J=0$ and $1$. Similary, Figure 5 presents
velocity profiles for the excited rotational levels $J=2$ and $3$.
Clearly, H$_2$ absorption near $-50$ km\,s$^{-1}$ 
is present at moderate strength in all of the lines shown,
whereas H$_2$ absorption from
the local Galactic disk gas is seen near $0$ km\,s$^{-1}$. We have
measured equivalent widths for the H$_2$ absorption in
the IV Arch (Table\,1) and have fitted the data to
curves of growth
for each individual rotational level, $J$, deriving
logarithmic column densities, log $N(J)$. For $J=0$,
the data fits best on a curve of growth with 
a very low Doppler parameter of $b=2.3^{+1.1}_{-0.5}$ km\,s$^{-1}$
(Figure 6), implying that the H$_2$ gas in the 
rotational ground state resides in a very confined 
region in the IV Arch that has a very low velocity 
dispersion. For $J=1,2$ and $3$, we find  
$b=3.5^{+1.4}_{-0.9}$ km\,s$^{-1}$, possibly indicating
that the excited molecular hydrogen gas is situated
in a less confined region. The different $b$ values
may reflect a core-envelope structure of the 
H$_2$, in which the more turbulent gas in the 
outer parts is rotationally excited, whereas
the inner, less turbulent part of the H$_2$
gas remains mostly in the rotational ground state.
However, the interpretation of the $J=0$ data points
in Figure 6 is difficult, since the lower $b$ value
for the rotational ground state is determined mainly
by the R(0) line with the lowest $f$-value (R(0),0-0
$\lambda 1108.128$; see Table\,1). If the equivalent
width of this line is overestimated by $1-2 \sigma$ 
(e.g., due to a noise peak or an unidentified intergalactic
absorption feature), the data points for $J=0$ would
also fit the curve of growth for the $J=1-3$ lines with 
$b=3.5$ km\,s$^{-1}$, and the H$_2$ column density for
$J=0$ would be lower. This seems somewhat 
unrealistic though, because the $N(0)/N(1)$ ratio (already
quite low with the two different $b$ values) would 
further decrease, and the H$_2$ column density 
distribution for $J=0$ and $1$ could not be fitted by an
excitation temperature that would be consistent
with the relatively low value for log $N(2)$ 
(see Figure 7). Thus, we 
adopt the two different $b$ values for $J=0$
and $J=1-3$ in the following discussions, but note that more 
precise data are required to improve the 
determination of $b$ for $J=0$.

Individual column densities, log $N(J)$, based on the
two different $b$ values for $J=0$ and $J=1-3$ given above, 
are listed in Table\,1.
The total H$_2$ column density, log $N($H$_2)$,
is $16.43\pm 0.14$ ($1\sigma$ error), and the fraction of hydrogen
in molecular form is
log $f=$ log $[2N($H$_2)/(N$(H\,{\sc i}$)+2N($H$_2))]=-3.3$,
based on the H\,{\sc i} 21cm data from Effelsberg.
We have derived excitation temperatures for the
H$_2$ gas by fitting the H$_2$ rotational level population
to a theoretical Boltzmann distribution (Figure 7). We
find an excitation temperature, $T_{\rm 01}$, of $141\pm20$ K
for $J=0$ and $1$, possibly 
representing an upper limit for the kinetic temperature of the gas in the
interior of the cloud. The temperature in the 
inner core of the cloud could be lower,
if part of the H$_2$ gas with $J=1$ resides in a more
extended region than the $J=0$ gas, as implied by the slightly higher 
$b$ value for $J=1$.
For $J=2,3$ and $4$ (upper limit) we find $T_{\rm 24}=885\pm270$ K.
This high excitation temperature probably reflects
several competing processes, such as UV pumping and H$_2$
formation pumping
(see Shull \& Beckwith 1982).
The column density ratios $N(3)/N(1)$, $N(4)/N(2)$, and 
$N(5)/N(3)$ can be used to disentangle the various excitation
mechanisms to a certain degree 
(e.g., Browing, Tumlinson \& Shull 2002), but such an
interpretation is difficult if information on the 
higher rotational levels is limited.
In view of the uncertainties for $N(2)$ and $N(3)$, and the fact that
no H$_2$ is detected for $J=4$ and $5$ we 
therefore refrain from a more detailed analysis
of the rotational excitation of the IVC H$_2$ gas toward PG\,1351+640.

The total H$_2$ column density in the 
IV Arch towards PG\,1351+640, log $N=16.43 \pm 0.14$,
is the highest that has been reported for intermediate-velocity 
clouds so far. This is not surprising, however, since the line
of sight to PG\,1351+640 passes through the IV Arch 
in a region with an
H\,{\sc i} column density of log $N=20.05$, which is substantially
higher than in other directions for which H$_2$ absorption
in IVCs was measured before (e.g., Gringel et al.\,2000; Richter
et al.\,2001a, 2001b). 

From a measurement of several atomic absorption
lines of Fe\,{\sc ii}, Si\,{\sc ii}, and P\,{\sc ii} in the 
PG\,1351+640 spectrum
we find atomic column densities of log $N$(Fe\,{\sc ii})\,$\approx 15.1$,
log $N$(Si\,{\sc ii})\,$\approx 15.2$, and 
log $N$(P\,{\sc ii})\,$\approx 13.6$ for the IV\,19 component at $-50$ km\,s$^{-1}$.
Together with the H\,{\sc i} column density of log $N=20.05$
from the 21cm data we derive logarithmic gas-phase abundances,
[$X$/H] = log($N_X/N_{\rm H\,{\sc I}})- $log($X$/H)$_{\odot}$, 
of [Fe/H]\,$\approx -0.4$, [Si/H]\,$\approx -0.4$, and 
[P/H]\,$\approx0.0$ (the values for log($X$/H)$_{\odot}$ are 
from Anders \& Grevesse 1989; Grevesse \& Noels 1993).
Thus, the FUSE and the H\,{\sc i} 21cm data imply that IV\,19
has a nearly solar abundance (as sampled by P), similar to what
is found for the IV Arch in the direction of PG\,1259+593
(Richter et al.\,2001b). Fe and Si appear to be  
under-abundant, possibly due to depletion into
dust grains, on whose surface the H$_2$ formation
is taking place. It is important to note that
this abundance determination is based on a beam-smeared
H\,{\sc i} column density. If substantial sub-structure
exists within the Effelsberg beam ($\sim 9$ arcmin), as 
possibly implied by the existence of H$_2$,
the abundances cited above might be incorrect.
Interestingly, the $b$-value for
the atomic absorption is rather high ($\geq 20$
km\,s$^{-1}$), indicating that the H$_2$ indeed
resides only in the inner-most region of IV\,19,
surrounded by a much larger envelope 
(with a higher velocity dispersion) in which the
bulk of the neutral gas is situated. 

No H$_2$ absorption
is seen in the Complex C component at $-155$ km\,s$^{-1}$,
giving further observational evidence that this metal-poor
HVC is not able to form and maintain diffuse molecular
hydrogen (see Richter et al.\,2001c). We estimate an
upper limit of log $N \leq 14.45$ ($3\sigma$) for the 
total H$_2$ column density in Complex C
toward PG\,1351+640.
This limit is consistent with previous H$_2$ measurements
in Complex C (Murphy et al.\,2000; Richter et al.\,2001c; Collins, Shull, \& Giroux 
2002).   

\section{Distribution of H$_2$ in IVCs}

Only a few cases of H$_2$ absorption in IVCs have
been reported in the literature (Gringel et al.\,2000;
Richter et al.\,2001a, 2001b; Bluhm et al.\,2001),
and thus very little is known about the overall
extent of molecular gas in intermediate-velocity
clouds. Previous detections include the IV Arch, the
LLIV Arch, and the IVC in front of the LMC.
Column densities and molecular gas fractions
derived from these studies are lower than those
found in the IV Arch towards PG\,1351+640,
with logarithmic column densities
between $14.1$ to $15.7$ and
molecular hydrogen fractions, log $f$, between
$-5.4$ and $-2.2$. All H$_2$ measurements
so far imply a diffuse molecular gas phase in IVCs.

To learn more about the distribution of diffuse molecular hydrogen
in IVCs and the physical conditions
of the gas in which the H$_2$ resides, we have searched for
H$_2$ absorption with FUSE
in intermediate-velocity clouds towards
$\sim 200$ extragalactic backgound sources,
such as quasars and active galactic nuclei (AGNs),
as well as two distant halo stars. This data
is available from the survey of O\,{\sc vi} absorption in the
thick disk of the Milky Way, as described by Savage et al.\,(2002)
and  Wakker et al.\,(2002).
It is important to note that this data set
is inhomogeneous with respect to sky-coverage, signal-to-noise (S/N),
and overall data quality. Most of the background sources
are located in the northern Galactic sky, giving access to the
large northern IVC complexes, such as the
IV Arch and the LLIV Arch.
As described in \S2, only a small fraction of the
suitable backgound sources observed with FUSE can
actually be
used to study interstellar absorption in IVCs due
to the flux limitations of these objects.
From the available sight lines,
56 have FUSE data with a S/N sufficient
to study interstellar H$_2$ absorption,
and are known to pass
intermediate-velocity H\,{\sc i} halo gas at
$|b|>25$, $|v_{\rm LSR}| \ge 25$ km\,s$^{-1}$,
and log $N$(H\,{\sc i})$\geq 19.0$.
44 of these are located in the northern 
Galactic sky, 12 in the south.

Table 2 lists the 56 background sources
and their coordinates together
with the IVC velocities
and IVC H\,{\sc i} column densities.
IVC identifications have been adopted from
Wakker (2001).
H\,{\sc i} 21cm emission line data for these directions
is available from the
Leiden-Dwingeloo Survey (Hartmann \& Burton 1997),
the Effelsberg 100m telescope (Wakker et al.\,2001),
the Green Bank 140\,ft telescope (Murphy et al.\,1996),
and the Villa Elisa observatory (Arnal et al.\,2000).
\footnote{The $3\sigma$ detection limits for $N$(H\,{\sc i}) from
these surveys typically vary between $3$ and $10
\times 10^{18}$ cm$^{-2}$.}
For more details on the H\,{\sc i} spectra see Wakker
et al.\,(2001).
Note that there are some sight-lines
that have multiple IVCs in the direction
of the background source. Here, we consider only
multiple IVCs that can be resolved with FUSE
(i.e., components that are separated in velocity
by more than $20-25$ km\,s$^{-1}$).
Two sightlines
(PKS\,0558-504 and NGC\,1705; formally included in Table 2)
show H\,{\sc i} 21cm IVC
emission in three components that have no counterparts in any atomic
absorption line in the FUSE spectrum; these emission features 
are probably caused by H\,{\sc i}
beam smearing effects and/or radio base-line calibration
errors.
We do not consider these cases any further.
In summary, the FUSE data set consists of
61 convincingly detected H\,{\sc i} IVC components at
log $N$(H\,{\sc i})$\geq 19.0$
that are also detected in FUV metal-line absorption.
We consider H$_2$ absorption in these
IVC components in the following paragraphs.

A S/N of $\sim 9$ (per resolution element) is typically required
to identify IVC H$_2$ absorption in individual lines
at equivalent widths of $\geq 50$ m\AA. Generally,
a clear identification is easier, if the IVC
component is well separated from the local
Galactic gas (i.e., at velocities $\geq 50$ km\,s$^{-1}$).
For one case (Mrk\,9; see Table\,2), the local Galactic
H$_2$ absorption is so strong, that it totally overlaps
possible $J=0,1$ H$_2$ absorption at intermediate
velocities, so little can be said about IVC H$_2$ absorption
in this particular case.
Together with the varying strength of local Galactic H$_2$ absorption
and the resulting difficulty of separating intermediate-velocity
components at varying S/N from the local Galactic component,
the detection limits for H$_2$ absorption at intermediate velocities
vary typically between $20$ and $80$ m\AA.
Note that the S/N and the $3\sigma$ detection limit may vary
considerably within a spectrum due to flux variations
and instrumental effects (e.g., fixed-pattern noise, blending
problems, varying exposure times for different
detector segments).
Out of 29 IVC components in FUSE spectra that
have relatively good S/N (typically $\sim 9$ and higher)
and a sufficiently
resolved IVC component, H$_2$ absorption in individual lines
at intermediate velocities is clearly detected
in 14 cases,
including the sight lines to PG\,0804+761
(sampling the LLIV Arch) and PG\,1259+593
(sampling the IV Arch) for which IVC H$_2$
absorption was reported earlier in individual
sight-line analyses (Richter et al.\,2001a; 2001c).
H$_2$ absorption in
the IV Arch has also been reported for the line of
sight toward HD\,93521, using ORFEUS data
(Gringel et al.\,2000); this sight-line
has been included in Table 2. We here do not
consider H$_2$ absorption that has been
detected in the low column
density IVC gas (log $N$(H\,{\sc i})$<19.0$)
toward the LMC (Bluhm et al.\,2001). This
general direction will be studied in detail in a
different paper (Richter, Sembach \& Howk 2002).
H$_2$ equivalent widths for
the 10 previously unpublished IVC components are given in Table 3.
We list six equivalent widths per IVC component, representing
the most pronounced IVC H$_2$ features for each
case. The number of H$_2$ IVC features detected
typically varies between 10 and 20 per
component and are mostly
$3-5\sigma$ detections. 
In Table 4 we list H$_2$ column densities,
log $N(J)$ and log $N($H$_2)_{\rm total}$, and $b$-values that
emerge from a fit of the data points in
a fashion similar to that described for PG\,1351+640
(see \S3 and Figure 6).
With the exception of NGC\,4151, the $1\sigma$ uncertainties
for the total H$_2$ column densities (see last column
of Table 4) are higher than for PG\,1351+640. 
The main contribution 
to the error is the determination of $b$, which turns out to be
problematic for many of the IVC H$_2$ components where the S/N
is low and/or the number of H$_2$ lines for
certain values of $J$ is limited due to blending problems.     
More accurate data are required to improve
the accuracy of the determinations of log $N($H$_2)$. 
In all 14 cases,
the fraction of hydrogen in molecular form
is small (log $f<-3$).

For the sight line
towards 3C\,273, Sembach et al.\,(2001b) have reported
H$_2$ absorption in FUSE data at positive velocities
around $+16$ km\,s$^{-1}$,
offset from the main local Galactic absorption (near
$-15$ km\,s$^{-1}$, see
their paper for details). The H\,{\sc i} 21cm emission
line data from the Greenbank 140ft telescope
(Murphy et al.\,1996) is quite complex and
shows three main H\,{\sc i}
components: a narrow emission feature near $-6$ km\,s$^{-1}$
and two broad components
located near $-20$ km\,s$^{-1}$ and $+25$ km\,s$^{-1}$.
We have re-analyzed the FUSE data of 3C\,273 using
the newest available version of the FUSE pipeline
(v2.0.5) in order to clarify the H$_2$ velocity structure
of the individual Galactic absorption components. The new
reduction (which provides a significantly improved velocity
calibration of the FUSE data) confirms that
the main H$_2$ absorption is offset in velocity from the main
local absorber near $-15$ km\,s$^{-1}$. It is now clear
that the H$_2$ absorption is related to the
H\,{\sc i} IVC gas at positive velocities around $+25$ km\,s$^{-1}$.
We therefore have included the line of sight towards
3C\,273 as a positive detection of H$_2$ at IVC velocities
in our sample (see Table 2). The  H$_2$
column density listed in Table 2 
has been adopted from Sembach et al.\,(2001b),
since (except for the velocity calibration) the shape of the H$_2$ absorption
spectrum produced by the v2.0.5 pipeline is identical with the spectrum
from the earlier pipeline version used by Sembach et al.\,(2001b). 
 
Figure 8 shows typical
H$_2$ absorption profiles for the cases
in which IVC H$_2$ absorption is newly
detected (plots for PG\,1259+593 are
presented in Richter et al.\,2001c).
The IVC components and the local
Galactic component are marked for
each profile with dotted lines. Total
H$_2$ column densities are also listed
in Table 2, together with the molecular
hydrogen fractions, log $f$,
based on the H\,{\sc i} column densities
from the 21cm observations listed in the
seventh column of Table 2.

For the remaining IVC components in
FUSE data with low S/N we find evidence
for H$_2$ absorption in IVC gas in additional
17 cases. We have produced co-added composite velocity profiles
for each spectrum including several H$_2$ lines
from various transitions in order to improve
the S/N and to study the general
velocity extent of the H$_2$ absorption.
We claim a tentative
detection of H$_2$ in IVC gas (labeled
as ``possible'' detections in Table\,2),
when the composite H$_2$ absorption profile
shows $\geq3\sigma$ evidence for an
absorption component at intermediate
velocities that would match the
H\,{\sc i} 21cm emission line data.
However, more accurate data are required to
confirm the presence of IVC H$_2$
for these cases. $3\sigma$ upper limits
for the total H$_2$ column densities
and the molecular hydrogen fractions
are listed in Table 2. The column density limits
are based on the $3\sigma$ equivalent width
limits of various blend-free H$_2$
transitions from the rotational states
$J=0-4$ that have been measured for each
of the 17 IVC components. The total H$_2$ column density
limit for each IVC component has been
calculated using a curve-of-growth
technique, assuming that $b\geq3$ km\,s$^{-1}$
for all $N(J)$.
Figure 9 shows three examples of
co-added composite H$_2$ velocity profiles
for NGC\,7714, Mrk\,357, and Mrk\,618.
At least three H$_2$ lines have been co-added
together per sight line, suggesting that
H$_2$ absorption is possibly present at IVC
velocities along each sight line.

We find no evidence for H$_2$ absorption
at IVC velocities in 30 of the 61 IVC components
measured with FUSE. Upper limits for the
H$_2$ column densities and the molecular hydrogen
fractions (derived using the same procedure
as for the tentative cases) are given in Table 2.
It is possible, however, that diffuse molecular
hydrogen is present below the
detection limit. This is
particulary likely for spectra that have low
S/N (see last column of Table 2). 
A good example for this is the
FUSE spectrum of PG\,1259+593:
H$_2$ in the IV Arch is detected at a very
low column density of log $N=14.10$
at a very high S/N of $\sim 30$. 
If the S/N were lower by
50 percent, H$_2$ in the IV Arch would be
totally invisible along this sight line.
Therefore, it is likely that with a better
overall data quality the fraction
of sight lines showing H$_2$ in IVC gas would
increase substantially.
For the IVC component at $+62$ km\,s$^{-1}$ towards NGC\,3783,
Sembach et al.\,(2001a) give an upper limit for the H$_2$
column density of log $N($H$_2)\leq15.00$, assuming that 
$b\geq 5$ km\,s$^{-1}$. Using our method
as described above we allow $b$-values as low as
$3$ km\,s$^{-1}$, so we find for the same component
a more conservative upper
limit of log $N($H$_2)\leq15.84$ for the same FUSE data.
For the second IVC component along this sight line at
$+34$ km\,s$^{-1}$ (see Table\,2), we find an even
higher limit of log $N($H$_2)\leq16.52$
due to the fact that this component is partly blended by
the local Galactic absorption near zero velocities.

Summarizing, we have investigated molecular
hydrogen absorption with FUSE in 61 H\,{\sc i} IVC components.
Significant H$_2$ absorption is detected and
measured in 14 cases. An additional 17 components
show evidence for H$_2$ absorption, but more
precise data is required to confirm the
presence of H$_2$. No evidence for H$_2$ is
seen for 30 IVC components, but H$_2$ might
be present at levels below the individual
detection limits.

\section{Comparison with H\,{\sc i} Observations}

Figure 10 shows the location of the 56 IVC sight lines
(as listed in Table 2) in the northern and southern Galactic sky in polar
projection, plotted over the IVC H\,{\sc i} 21cm distribution
at intermediate velocities between $v_{\rm LSR}=-30$ to
$-90$ km\,s$^{-1}$ (from data of the Leiden-Dwingeloo survey,
Hartmann \& Burton 1997).

Filled boxes show the directions, in which H$_2$ is detected
at IVC velocities, filled triangles mark
the low-S/N cases where IVC H$_2$ is possibly detected, and
open circles indicate the
non-detections.
IVC H$_2$ absorption has a distribution 
as widespread as that of the H\,{\sc i} 21cm emission, 
although there are some
sight lines, where H\,{\sc i} emission is seen at log $N \geq 19.3$
but no H$_2$ absorption is detected, even at
high S/N (e.g., Mrk\,209). 

In Figure 11 we have plotted the fraction of hydrogen in
molecular form, log $f=$\,log $[2N($H$_2)/(N$(H\,{\sc i}$)+2N($H$_2))]$
against the logarithmic H\,{\sc i} column density (from the 21cm data)
for the 61 IVC components listed in Table 2.
In this plot, the H$_2$ detections are plotted as filled
circles, whereas the possible detections and the non-detections
have open symbols (open triangles for the possible detections, open
circles for the non-detections). Note that all
open symbols plotted in Figure 11 therefore represent upper limits
for log $f$, while the filled circles show measured
molecular hydrogen fractions.
The very inhomogeneous data quality of the sample clearly
hampers a statistical analysis of the H$_2$ fraction in
IVCs as a function of the total gas column density.
Data points for log $f$ vary between $-1.4$ (upper limit)
and $-5.3$ (measured value); thus they
span a range over $\sim 4$ orders of magnitude.
For the positive H$_2$ detections there is no visible correlation
between log $f$ and log $N($H\,{\sc i}$)$ and the scatter
for the data points in log $f$ is substantial. One could
expect a correlation in the way that the molecular
hydrogen fraction increases with the total gas column
along the sight line because of H$_2$ line self-shielding.
The fact that no correlation is seen implies that 
the H$_2$ absorption comes from low-column density
H$_2$ aborbers without self shielding, while the scatter
could be due to the varying number of H$_2$ absorbers per sight line.
However, the observed scatter may also indicate
that a) the H\,{\sc i} column densities measured in 21cm do not 
represent accurate values
for the pencil beam sight lines because of beam-smearing effects, 
b) some fraction
of the neutral gas is not physically
related to the molecular material along the various lines of sight.

It is interesting that all of the
detected H$_2$ IVC components have low molecular hydrogen
fractions with log $f\leq -3.3$. This suggests that for the
many upper limits with log $f>-3.3$ there is a substantial
number of IVC components that have molecular material at
H$_2$ fractions that are undetectable at the current S/N
of the FUSE data.
From the $29$ data points with log $f\leq-3.0$ there are
$14$ positive detections, $4$ possible detections, and $11$
non-detections. If this distribution is representative,
H$_2$ at low molecular fractions may exist in over 50 percent
of all IVC components that have H\,{\sc i} column densities
log $N($H\,{\sc i}$)\geq 19.00$ and thus is a surprisingly
widespread constituent of the gas in the lower Galactic halo.
For the $14$ positive IVC H$_2$ detections the 
mean logarithmic molecular hydrogen
fraction is $-4.3$. 

\section{Origin of the H$_2$}

To understand the presence of molecular hydrogen
in IVCs and its role for
the interstellar gas in the lower Milky Way halo
it is important to consider
possible formation sites of the H$_2$.

Although it may be possible
that molecular cloud complexes are pushed
out of the disk of the Milky Way
by energetic events (see Leass et al.\,2002,
in preparation),
it appears very unlikely (and there is no
observational evidence) that entire molecular clouds
are transported several kiloparsecs above the
Milky Way disk, and that the observed H$_2$ represents
a diffuse remnant from dense molecular clouds that
have been elevated from the disk into the halo.
Two noticable exceptions
may be the Draco cloud and IV\,21 (see above), which
have been detected in CO emission, but these
IVCs are located in the disk-halo interface ($z\leq500$ pc)
and they may well be ``normal'' molecular clouds at
exceptionally large $z$-heights.
Also, the presence of dust and heavy elements in
halo clouds that contain H$_2$ implies that
the observed molecular gas phase is not sampling
any primordial molecular gas clumps in the halo
that have been proposed in the literature as
candidates for baryonic dark matter (e.g., de Paolis
et al.\,1995, Kalberla, Shchekinov \& Dettmar 1999). The roughly solar
abundances that are found in IVCs instead suggests that
this gas represents Galactic disk material that is
somehow circulating from the disk into the halo and back, possibly
by way of a Galactic Fountain (Shapiro \& Field 1976).
In this model, gas ejected out of the disk
by way of supernova explosions cools and falls back
onto the Milky Way disk.
Similar processes are probably also seen in other 
galaxies. For instance, the many extraplanar dust
filaments observed in edge-on galaxies such as NGC\,891 
(Howk \& Savage 1997) suggest that Fountain-type processes
may be a rather common phenomenon in spiral galaxies.

The existence of diffuse H$_2$ at relatively
low molecular hydrogen fractions implies that
the H$_2$ line self shielding (see, e.g., 
Draine \& Bertoldi 1996) is probably not
very efficient in such gas. Without the
self shielding, however, the photo-dissociation
time scales are quite short ($\sim 10^3$ years),
and no diffuse molecular material should be
detectable, unless the dissociation is counteracted
by local H$_2$ {\it formation}. Following this logic,
the widespread existence of H$_2$ in IVCs strongly
suggests that molecule formation takes place {\it in situ}
in the most dense regions of IVCs. It has been proposed
that halo clouds consist of a multiphase structure
(Wakker \& Schwarz 1991; Wolfire et al.\,1995) with
cold, dense cores (``Cold Neutral Medium'', CNM)
embedded in a more tenuous, warm medium 
(``Warm Neutral Medium'', WNM). 
Most likley, 
the diffuse molecular hydrogen forms in small,
dense filaments during the cooling and
fragmentation phase of a Galactic Fountain, and
traces the CNM in 
the intermediate-velocity clouds.
Houck \& Bregman (1990) have developed a model
for low-temperature Fountains with
a cooling timescale, $t_{\rm cool}$, of
$\sim 3 \times 10^7$ yr, and a time scale for
newly formed clouds to return to the disk,
$t_{\rm ret}$, on the order of $5 \times 10^7$ yr.
This gives the CNM sufficient time to form H$_2$
at moderate densities to the fractional level 
that is observed (see \S7.2). 
Since dust is available in IVCs,
we assume that the H$_2$ formation takes place on the
surfaces of dust grains, the most efficient way to
form molecular hydrogen.  
One might argue that the grain
formation rate in the halo must be significantly lower than in
the disk due to the highly energetic processes operating that might destroy
or at least modify the properties of the dust grains.
The results of Sembach \& Savage (1996) in fact 
indicate that the surfaces of grains
in halo clouds are probably different from those in diffuse 
disk clouds, but the implications for the halo H$_2$ grain
formation rate are unclear.
Franco et al.\,(1991) have presented the 
idea of a ``soft'' Galactic Fountain.
They propose that
dust grains might be transported into the halo due to
radiation pressure of starlight
near regions of intense star formation. It is shown that this
``photolevitation'' may push small diffuse clouds above the
disk, whereas massive molecular clouds would not be affected by the
momentum transfer from the radiation field.
Such a process
may only mildly modify the dust properties and could well
explain the widespread presence of dust and molecular
hydrogen in the lower Galactic halo. Franco et al. find
that the photolevitation effect can maintain a neutral
gas column density of $\sim 10^{20}$ cm$^{-2}$ above the disk,
similar to what is found for most of the IVCs.

\section{Physical Conditions}

The overall ubiquity of H$_2$ in IVCs is
indicating that these clouds must contain
substructure,  
providing small, relatively dense cloudlets and
filaments that allow H$_2$ to form at a 
rapid enough rate to compensate for
the destruction by UV photo-dissociation.
In the following, we will investigate
densities and typical dimensions
of the structures in which the H$_2$ may reside by
discussing the processes and parameters that
balance the formation and dissociation of
molecular hydrogen in Galactic halo clouds.

Since dust is present in IVCs
(see Richter et al.\,2001a, 2001b), we
consider H$_2$ dust grain formation as the
dominating H$_2$ formation process in IVCs
and ignore any
gas phase reactions
(see Black 1978).

\subsection{H$_2$ Formation-Dissociation Equilibrium}

In an H$_2$ formation-dissociation equilibrium
in transparent clouds,
the H$_2$ volume
density, $n$(H$_2$), is related to
the H\,{\sc i} volume density, $n$(H\,{\sc i}), via
\begin{equation}
\frac{n({\rm H\,I})}{n({\rm H}_2)} =
\frac{\langle k \rangle \beta_0}{Rn_{\rm H}},
\end{equation}
\\
\noindent
where at low molecular hydrogen fractions $n_{\rm H} = n$(H\,{\sc i})
$+ 2n$(H$_2$)$ \approx n$(H\,{\sc i})
in cm$^{-3}$ (Spitzer 1978, hereafter S78).
In this equation, $\langle k \rangle \approx  0.11$ is the
probability that the molecule is dissociated after photo absorption,
$\beta_0$ is the photo-absorption rate per second, and
$R$ is the H$_2$ formation rate in units cm$^{3}$\,s$^{-1}$.

The H$_2$ photo-absorption rate $\beta_0$ in the Galactic halo depends
on the mean FUV radiation field at a height $z$ above the
Galactic plane. Principally, the FUV radiation field is expected
to be lower in the Milky Way halo in comparison to the disk,
and various attempts
have been made to find scaling relations for the radiation field
in the halo as function of the $z$-height
(e.g., Wolfire et al.\,1995). These studies show that only a
fraction 
of the ultraviolet photons that are produced
in the Galactic disk (mainly by OB stars) escape
into the halo, mainly due to the extinction by dust grains.
For the lower Galactic halo,  
the $z$-dependence of UV field in the halo is small due
to the large solid angle of the Galactic plane.
Using equation (4) from Wolfire et al.\,(1995), the 
FUV radiation field at a $z$-height of 1 kpc 
is reduced by a factor of $\sim 2$ in comparison
to the midplane intensity.
The H$_2$ photo-absorption
rate in the halo is expected to scale directly with
the halo UV field and thus is reduced by a similar
factor. In the disk, the mean value
for $\beta_0$ is $5.0\times 10^{-10}$ s$^{-1}$ (S78).
We therefore assume that $\beta_{\rm 0,halo} \approx
0.5\,\beta_{\rm 0,disk} = 2.5\times 10^{-10}$ s$^{-1}$ at
the outer edge of an IVC.
H$_2$ line self-shielding reduces the UV flux in
the interior of a cloud once the H$_2$ column density
in a single absorbing component exceeds a critical value.
Following the approximation given by Draine \& Bertoldi (1996),
the H$_2$ photo-absorption rate then is reduced by the factor
$S=(N($H$_2)/10^{14}$ cm$^{-2})$ for H$_2$ column densites
log $N($H$_2)\geq 14$. 
Whether or not the H$_2$ self-shielding is important
for IVCs is unclear. The total H$_2$ column 
densities measured in IVCs are log $N($H$_2)=14.1-16.4$, 
implying that for single-component absorbers self shielding
may reduce the photo-absorption rate in the innermost 
region down to a level of $\sim 4.0  
\times 10^{-12}$ s$^{-1}$. However,
the relatively high $b$ values measured for most of these clouds
(Table 4) imply that there are several absorbing CNM structures within an IVC 
contributing to the total H$_2$ column density
(see also discussion in \S5), and H$_2$ self-shielding 
probably plays therefore only
a minor role. We will discuss situations with and without
self-shielding at the end of this section. 

Because we assume
that all of the H$_2$ in IVCs forms on the surfaces of dust grains,
the formation rate parameter $R$ given in equation (1) denotes the 
rate per neutral H atom to form H$_2$ molecules by collisions with
dust grains.
$R$ is a function of the gas and dust temperature, the sticking
coefficient of H atoms on the grain surface,
the mean
projected surface area of the dust grains, and the mean H volume
density (see S78). 
The value for $R$ in Galactic halo clouds is unknown, as there
is little knowledge about the properties of dust grains away from
the Galactic disk. Here, we
simply assume that $R_{\rm halo} \approx R_{\rm disk} = 
3 \times 10^{-17}$ cm$^3$\,s$^{-1}$ (S78) due to
the lack of better information. It is possible, however,
that $R_{\rm halo}$ differs significantly from the value
measured in the disk. 

One major problem that arises from equation (1) is that the
formation-dissociation equlibrium relates the volume densities
$n$ of H\,{\sc i} and H$_2$ with each other, while
column densities $N$(H\,{\sc i}) and $N$({H$_2$}) are
measured. It is often assumed that
$N$(H\,{\sc i})/$N$(H$_2$) $ \approx $ $n$(H\,{\sc i})/$n$(H$_2$)
(see S78), but this assumption might be incorrect in the case
of Galactic halo clouds because they likely have a pronounced multi-phase
structure and unresolved sub-components. 
It is expected that IVCs (as well as
HVCs) consist of large gaseous envelopes that contain
both ionized and atomic species, interacting with the
hot Galactic corona and residing in pressure equilibrium
with the hot coronal gas (Spitzer 1956; Sembach et al.\,2002). 
Most likely, the neutral and partly molecular cloud core that 
is associated with the CNM is embedded
in a low density, higher temperature layer representing 
the WNM, but it is
unknown in which phase the bulk of the gas is
residing. In any case, the measured $N$(H\,{\sc i}) represents the
{\it total} column density of neutral hydrogen along the
line of sight, and all gas phases that contain
neutral gas at whatever fraction will contribute to its
value. Thus, it appears not to be justified to simply compare
$N$(H\,{\sc i}) and $N$({H$_2$}) in IVCs, since the H$_2$
likely resides in regions much more confined than the H\,{\sc i},
and only a fraction of the total 
H\,{\sc i} is available for the formation of
hydrogen molecules.
The different gas phases (we here assume just two phases:
neutral-molecular and neutral-ionized) should be treated
separately, and a scaling factor for the H\,{\sc i}
column density in the IVCs is required. We define
$\phi$ as an H\,{\sc i} scaling factor $\le 1$,
so that

\begin{equation}
\left( \frac{n({\rm H\,I})}{n({\rm H}_2)} \right)_{\rm f} =
\phi \, \frac{N({\rm H\,I})}{N({\rm H}_2)},
\end{equation}
\\
where ``f'' stands for the dense filament
in the CNM in which the H$_2$ gas resides, and $N$(H\,{\sc i}) is
the total H\,{\sc i} column density of the IVC measured along the
line of sight ({\it not} including other H\,{\sc i}
velocity components).
Note that the molecular hydrogen fractions listed in Table 2
(tenth column) are derived in the standard way from the {\it total} IVC H\,{\sc i}
column densities given by the 21cm observations.

Using the values for $\beta_0$ and $R$ 
for the Milky Way halo (as defined above) and
equation (2), we find
for the hydrogen volume density in the region
of the IVC that contains the H$_2$:

\begin{equation}
n_{\rm H,f} \approx 9.2 \times 10^5 \, \frac{N({\rm H}_2)}{N({\rm H\,I})}
\phi^{-1} .
\end{equation}
\\
\noindent
Equation (3) allows us to determine the hydrogen volume density in the
partly molecular cloud core of an IVC, assuming that the observed
fraction of hydrogen in molecular
form represents its abundance in formation-dissociation equilibrium.

If we neglect H$_2$ line self-shielding and take 
$N$(H$_2$)/$N$(H\,{\sc i})$=2.5 \times 10^{-5}$, 
$\beta_{\rm 0,halo} = 2.5\times 10^{-10}$ s$^{-1}$, and
$\phi=0.5$
as typical values, we derive
a hydrogen volume density of $n_{\rm H,f} \approx 50$ cm$^{-3}$ by
using equation (3). If self-shielding is considered, the 
volume density is reduced by the factor $S$ and
could be as low as $\sim 1$ cm$^{-3}$. However, if the
CNM in IVCs is broken up into small column density filaments (see above),
volume densities of $n_{\rm H}=10-50$ cm$^{-3}$ 
appear to be more realistic. 
 
For the somewhat stronger absorption
in the IV Arch towards
PG\,1351+640 we have
$N$(H$_2$)/$N$(H\,{\sc i})$\approx 2.6 \times 10^{-4}$
(see \S3),  
so that with $\phi=0.5$
and without self-shielding
the hydrogen volume density comes out as high as
$n_{\rm H,f} \approx 500$ cm$^{-3}$. 
With self-shielding, we obtain
$S=(2.69\times10^{16}$cm$^{-2}/10^{14}$cm$^{-2})^{-0.75}\approx 0.015$
and $n_{\rm H,f} \approx 7$ cm$^{-3}$.
The actual value for $n_{\rm H,f}$ probably lies 
somewhere in between, but without knowing the
3D structure of the H$_2$ absorber it is difficult to estimate
a more realistic value for the self-shielding factor $S$.
However, if we assume that the mean thermal pressure in the CNM in these
clouds is similar to that in the disk ($P_{\rm CNM}/k =n_{\rm H}\,T=4000$ cm$^{-3}$ K;
Heiles 1997), the rotational excitation temperature $T_{01}=141$ K derived in
\S3 implies $n_{\rm H,f} \approx 30$ cm$^{-3}$.

The linear thickness of such a CNM/H$_2$ structure,
$D=\phi\,N$(H\,{\sc i})\,($n_{\rm H,f})^{-1}$,
is small, $\sim 0.1$ pc typically. 
Such a small filament is then probably
surrounded by a low-density WNM envelope  
that has a much larger extent.
Evidence for filamentary sub-structure of IVCs also
comes from high-resolution optical spectra
(Ryans, Sembach \& Keenan 1996; Welty et al.\,1999), in which 
multiple velocity components are detected in IVC gas separated only
by a few km\,s$^{-1}$. Because of the close 
proximity of the components in velocity space 
it appears likely that they are physically 
related to each other (i.e., they are sub-components
of the same IVC).
 
If IVCs are cooling and falling back onto the Milky Way
disk (as indicated by their radial velocities), the most
dense gas conglomerates may
be located relatively close ($z<1$ kpc) to the midplane,
i.e., in the disk-halo interface.
At this stage, some of the molecular filaments 
might have turned dense enough to allow 
the formation of much larger amounts of H$_2$,
and possibly carbon monoxide (CO). Interestingly,
there are two IVCs known to exist in the disk-halo
interface that are the {\it only} IVCs showing CO emission: 
the Draco cloud (Mebold et al.\,1985) and IV\,21 
(Weiss et al.\,1999; Heithausen et al.\,2001).
Possibly, these two clouds represent the
final stage of a Galactic-Fountain IVC before re-entering
the Milky Way disk.

\subsection{H$_2$ Small-Scale Structure}

We have shown at the end of the previous section
the H$_2$ observations in IVCs suggest the
presence of a large number of small, relatively
dense filaments in the lower Milky Way halo, 
assuming that H$_2$ grain formation
and photo-dissociation are equilibriated.
In their best-fitting Galactic Fountain model
(model 3), Houck \& Bregman (1990) suggest
a number of $\sim 9 \times 10^4$ halo clouds at a 
size of $\sim 2.7$ pc above a disk 
area of 1 kpc$^2$. Assuming 
the thickness of typical H$_2$ filaments
is only $\sim 0.1$ pc, the relatively high 
detection rate of H$_2$ absorption measured in IVCs 
in this study 
implies that the number of H$_2$ bearing 
filaments per kpc$^2$ column is several orders
of magnitude higher.

It is likely that the diffuse H$_2$ gas traces the
CNM in these clouds, and that this CNM is surprisingly widespread,
given the high detection rate of H$_2$. We do not know,
however, whether these structures already represent the
highest density regions in IVCs, or whether more
dense molecular clumps with higher molecular fractions exist.
H$_2$ absorption line measurements are naturally biased 
towards diffuse clouds with low molecular fractions and low densities 
due to their large volume filling factor. The chance to detect such a
diffuse H$_2$ absorber by way of absorption spectroscopy 
towards a limited number of background sources is much higher
than to find a small, dense clump that would have 
a small angular extent. It has been demonstrated over 
the last few years that the ISM consists of substantial
small-scale structure at AU scales (e.g., Faison et al.\,1998;
Lauroesch, Meyer \& Blades 2000), indicating apparently
ubiquitous tiny gaseous structures (``tiny-scale atomic structures'', TSAS)
with very high densities ($n_{\rm H}\sim10^3-10^5$cm$^{-3}$). 
The TSAS are probably filamentary and sheet like structures
embedded in the CNM (Heiles 1997), but very little is 
known about their physical properties.
Very recently, tiny-scale structure in the ISM has also been
found in CO emission (Heithausen 2002), showing that 
molecules may play an important role for the understanding 
of these structures.

AU scale structure is 
also present in IVCs, as demonstrated by Na\,{\sc i}
observations of Complex gp against the globular cluster M\,15 
(Meyer \& Lauroesch 1999). At such high volume 
densities, those tiny structures in the halo should contain
molecular hydrogen at a high fractional abundance,
since the dissociating FUV field in the halo is not intense and
the H$_2$ formation time scale 
($t_{\rm form}=(R_{\rm halo}\,n_{\rm H})^{-1}$)
becomes short at high densities. 
For $n_{\rm H}=300$ cm$^{-3}$ the linear thickness would
be 10 times smaller than for $n_{\rm H}=30$ cm$^{-3}$,
if the H\,{\sc i} column density stays constant.
However, the surface filling factor would be down 
by a factor of 100 (assuming
spherical geometry). Thus, it would require a larger sample
of FUSE sight lines to be able to detect and measure
H$_2$ absorption in these smallest structures, and to
investigate the relation between the TSAS and the more diffuse 
CNM phase in intermediate-velocity clouds.  
A possible candidate for molecular absorption 
related to TSAS could be the 
intermediate- and high-velocity H$_2$ gas 
towards LH\,58 in the Large Magellanic Cloud
(Richter, Sembach \& Howk 2002; Richter et al.\,1999).

\section{Discussion}

Our study reveals that diffuse molecular hydrogen
is a widespread, but not very abundant constituent
in intermediate-velocity clouds in the low Galactic halo.
The results suggest that the H$_2$ in IVCs resides
in a large number of small, relatively dense 
filaments, associated
with the CNM in these clouds.

The ubiquity of H$_2$ in IVCs underlines that 
molecular hydrogen is able to form
and survive in regions that have low H\,{\sc i}
column densities, tracing the most dense
regions in the diffuse interstellar medium.
A key point for the formation
and survival of H$_2$ in IVCs is that these
clouds contain heavy elements and dust grains,
on which H$_2$ formation proceeds very efficiently.
Moreover, the FUV radiation field in IVCs is relatively low, reducing
the UV photo-dissociation of H$_2$. That the presence of dust
and the absence of a strong UV field is so
important for the existence of diffuse 
interstellar H$_2$ becomes
immediately clear when comparing our findings
with observations of H$_2$ in other
environments. For the Magellanic Clouds,
Richter (2000) and Tumlinson et al.\,(2002)
find that the average molecular hydrogen
fraction is significantly lower than in
the Milky Way, mainly because of the
very strong UV radiation from the many
young OB stars in the Clouds, and 
due to the lower dust abundance.
In the Magellanic Stream, Sembach et al.\,(2001a)
and Richter et al.\,(2001c) find H$_2$ absorption
in two of two available sight lines
at column densites of log $N($H$_2)=16-17$.
The Magellanic Stream contains heavy elements
and dust (Lu et al.\,1998), thus
allowing H$_2$ grain formation. In addition,
the UV radiation field in
the Magellanic Stream is expected to be low, resulting in
a rather moderate H$_2$ photo-dissociation rate
(see Sembach et al.\,2001a).
However, a low UV radiation field alone is not
sufficient to allow diffuse H$_2$ to exist
at moderate densities, as demonstrated by the
many non-detections of H$_2$ in metal-poor,
intergalactic Ly\,$\alpha$ clouds
(e.g., Petitjean, Srianand \& Ledoux 2000), 
and in high-velocity cloud
Complex C.
Towards PG\,1259+593, Richter
et al.\,(2001c) have studied H$_2$ absorption
in Complex C at a relatively high H\,{\sc i}
column density of log $N($H\,{\sc i}$) \approx 20$.
No H$_2$ was found down to a level of
log $N($H$_2) \approx 14$, although newer
Westerbork H\,{\sc i} 21cm interferometer
data clearly shows, that this sight line
passes a very confined H\,{\sc i} clump.
Complex C is more than 6 kpc away from the
Galactic plane and thus the UV radiation
field in Complex C is expected to be at 
least as low as in 
IVCs.
A detailed heavy element abundance study
(Richter et al.\,2001b) shows
that Complex C has a $\sim 0.1$ solar
metallicity in this direction and
the abundance pattern implies that
there is no or only very little
dust present. Therefore, H$_2$ grain
formation is suppressed, and gas phase
formation processes (Black 1978) are obviously not
efficient enough to allow diffuse H$_2$ gas
to exist in Complex C, even at relatively
high H\,{\sc i} column densities.

The current H$_2$ absorption line data strongly suggest that
the interplay of dust abundance and UV radiation
field is of prime importance for the abundance
of diffuse molecular hydrogen in the low-redshift
Universe. Still, many of the 
H$_2$ formation/destruction aspects remain unknown,
such as the grain-formation
rates in environments with low metallicities and precise
UV photo-dissociation rates. 
As we have shown, IVCs in the Milky Way halo
represent an interesting case
for the study of H$_2$ in
diffuse solar-metallicity gas that is exposed
to a moderate UV radiation field. Its
widespread distribution shows that 
H$_2$ might be a rather robust constituent 
of the diffuse ISM, if the overall conditions
are appropriate.
High S/N FUSE observations are desirable to derive
precise H$_2$ column densities, molecular fractions,
excitation temperatures in Galactic halo clouds,
and to search for the
presence of high-density H$_2$ clumps in IVCs and HVCs.
Such measurements might be of 
great importance for understanding and
predicting the abundance
of diffuse H$_2$ in the more distant Universe,
where H$_2$ is very difficult to measure systematically
with absorption spectroscopy. 

\section{Summary}

We have used FUSE absorption line data to
study the molecular hydrogen (H$_2$) content of
intermediate-velocity clouds (IVCs) in the
lower halo of the Milky Way.\\
\\
\noindent
(1) From 61 H\,{\sc i} IVC components that are sampled
in a FUSE data set of 56 sight lines at
latitudes $|b|>25$, $|v_{\rm LSR}| \geq 25$ km\,s$^{-1}$,
and log $N$(H\,{\sc i})$\geq 19.0$, we have detected
and analyzed molecular hydrogen absorption in 14 cases in
FUSE data with sufficient S/N ($\sim$ 9 per resolution 
element, and higher).
For an additional 17 IVC components with lower
S/N we also find
evidence for IVC H$_2$ absorption,
but cannot claim a firm detection.
Thirty IVC components show no evidence for the presence of H$_2$.\\
\\
\noindent
(2) We have measured H$_2$ column densities, 
log $N($H$_2)$, and molecular hydrogen 
fractions, log $f$, for the 14 positive IVC H$_2$ detections, and
upper limits for log $N($H$_2)$ and log $f$ for the remaining 47
components. For the positive IVC H$_2$ detections, log $f$ varies
between $-3.3$ and $-5.3$, with a mean of $-4.3$, 
implying a mostly atomic gas phase. 
Only 29 IVC components are sampled with FUSE data at a sufficient S/N 
to measure log $f$ in the range $\leq -3.0$ ($14$ detections, $4$
possible detections, and $11$ non-detections). For the rest of the 
data, IVC H$_2$ might also be present at low molecular fractions, but 
cannot be detected at the current data quality. 
Our study reveals that intermediate-velocity clouds in the
halo of the Milky Way contain a ubiquitous, diffuse
molecular hydrogen component.\\
\\
\noindent
(3) We have performed a detailed analysis of 
H$_2$ absorption 
in the Intermediate-Velocity
Arch (core IV\,19)
in the direction of the quasar PG\,1351+640. This sight line
has the most pronounced case of H$_2$ absorption
in IVCs found so far.
We find a total H$_2$ column density 
of log $N=16.43\pm0.14$, and a molecular hydrogen
fraction, log $f$, of $-3.3$. The low 
$b$-values ($b=2.3-3.5$ km\,s$^{-1}$) suggest that
the IVC H$_2$ gas resides in a very confined clump in
this direction. Two excitation temperatures are required
to describe the rotational excitation of the
H$_2$ gas: for $J=0$ and $1$ we derive $T_{\rm 01}=
141\pm20$ K, possibly representing an upper limit for 
the kinetic temperature
of the gas. For $J=2-4$ we find $T_{\rm 24}=885\pm270$ K.\\
\\
\noindent
(4) We have discussed the physical conditions 
in IVCs that are required to describe the distribution
and properties of the observed H$_2$. 
The solar metallicities of IVCs, as well as their
radial velocities, imply that most of
these clouds represent the return flow
of a Galactic Fountain, and that the
H$_2$ is forming 
on the surface
of dust grains
in the halo during the
cooling and fragmentation process.
If the H$_2$ 
stays in formation-dissociation equilibrium, the
molecular material must reside in a large
number of small ($D\sim0.1$ pc),
rather dense ($n_{\rm H} \sim 30$ cm$^{-3}$)
filaments. Most likely, the H$_2$ gas traces
the Cold Neutral Medium in these clouds. 
We speculate that more dense regions with higher
molecular fractions could exist,
associated  with Tiny Scale Atomic Structures.
These would be difficult to be detected in H$_2$
absorption due to their small size.\\
\\
\noindent
(5) H$_2$ measurements in
intermediate- and high-velocity clouds, in local
Galactic gas, in the Magellanic Clouds, and in
intergalactic Ly\,$\alpha$ absorbers 
indicate that the interplay between dust 
and the UV radiation field is crucial for the
abundance and distribution of diffuse
H$_2$ in the low-redshift Universe. 
If dust is present and the FUV radiation
field is moderate, H$_2$ appears to be a widespread
constituent of the diffuse interstellar
medium.

\acknowledgments

This work is based on data obtained for the
the Guaranteed Time Team by the NASA-CNES-CSA FUSE
mission operated by the Johns Hopkins University.
Financial support has been provided by NASA
contract NAS5-32985. PR is supported
by the {\it Deutsche Forschungsgemeinschaft.}
We thank the referee for helpful comments.

\clearpage
\newpage
Fig. 1.  H\,{\sc i} 21cm IVC sky (Aitoff projection) from the Leiden-Dwingeloo sky survey (LDS;
Hartmann \& Burton 1997) for LSR velocities between $-30$ and $-90$ km\,s$^{-1}$
and H\,{\sc i} column densities between $2$ and $70 \times 10^{19}$ cm$^{-2}$, 
plotted in Galactic coordinates ($l,b$) and centered on $l=120$.
The contours represent column densities of $2$ and $10 \times 10^{19}$ cm$^{-2}$.
For the identifications of the individual IVC complexes see Wakker (2001).
Note that the LDS data does not fully cover the sky for $b\leq+30$ in the
ranges $0\leq l \leq +15$ and $230 \leq l \leq 360$ (see Hartmann \& Burton 1997
for details).
\\
\\

Fig. 2. Continuum-normalized atomic absorption profiles of the FUSE
spectrum of PG\,1351+640 are plotted on a LSR radial
velocity scale, and compared to H\,{\sc i}
21cm emission line data from the Effelsberg 100m radio-telescope
(see Wakker et al.\,2001). Three major components are present
at $v_{\rm LSR} \approx 0, -50$ and $-155$ km\,s$^{-1}$, representing
gas from the local Milky Way, the IV Arch (core IV\,19, strongest component), 
and high-velocity cloud
Complex C, respectively. Complex C absorption is present 
in the Si\,{\sc ii} $\lambda 1020.7$ and Fe\,{\sc ii} $\lambda 1122.0$ lines,
but not in P\,{\sc ii} $\lambda 1152.8$. The H\,{\sc i}
emission line data contains substructure in each of the three
components, such as another IVC component at $-74$ km\,s$^{-1}$
(core IV\,9), which is also present in the atomic absorption 
profiles.
\\
\\

Fig. 3. A portion of the FUSE spectrum of PG\,1351+640 in the wavelength range
between 1076.5 and 1079.5 \AA, sampling the H$_2$ $2-0$ 
Lyman band. The individual H$_2$ lines present in the data are
labeled above the spectrum. The data plotted here shows H$_2$ absorption from
rotational levels $J=0,1$ and $2$ in two components at $0$ and
$-50$ km\,s$^{-1}$, representing H$_2$ gas in the local Milky Way disk
and the IV Arch (IV\,19), respectively. The solid gray line shows 
a two-component Gaussian fit of the spectrum that
reproduces the absorption pattern seen in the FUSE data.     
\\
\\

Fig. 4. Continuum-normalized H$_2$ absorption profiles in the FUSE
spectrum of PG\,1351+640 for the rotational
levels $J=0$ and $1$ are plotted. H$_2$ absorption is
present in the local Milky Way component at zero velocities
and in the IV Arch (IV\,19) component at $-50$ km\,s$^{-1}$.
H$_2$ lines are labeled in the upper left corner of each 
box (with `L' for the Lyman band and `W' for the Werner band).
Lines from other transitions or species are marked with `b'. 
\\
\\

Fig. 5. Same as Fig.\,3, but for rotational levels $J=2$ and $3$.
\\
\\

Fig. 6. Empirical curve of growth for H$_2$ absorption in the IV Arch (IV\,19)
toward PG\,1351+640 for the rotational levels $J=0-3$, as
labeled in the lower right corner.
Two very low $b$-values ($b=2.3^{+1.1}_{-0.5}$ km\,s$^{-1}$ for
$J=0$ and $3.5^{+1.4}_{-0.9}$ km\,s$^{-1}$ for $J=1-3$) are required to fit
the data in the log (W$_{\lambda}$/$\lambda$)-log ($Nf\lambda$) space,
suggesting that the H$_2$ ground state absorption occurs in a very
confined cloud core, whereas absorption from the excited levels
arises from the surrounding gas that has a slightly higher
velocity dispersion. A typical error bar that represents the
$1\sigma$ uncertainty in the H$_2$ equivalent widths (see also Table 1) is
shown in the lower left corner.
\\
\\

Fig. 7. H$_2$ rotational excitation in the IV Arch (IV\,19) toward PG\,1351+640.
The logarithmic column densities of each rotational
state, $N_J$, divided by the quantum mechanical statistical
weight for each state, $g_J$, are plotted 
against the excitation energy, $E_J$.
The dashed lines represent fits from a theoretical Boltzmann
distribution. The rotational ground states, $J=0$ and $1$, fit
on a line that is equivalent to a Boltzmann temperature of
$T_{01}=141\pm20$ K, possibly representing an upper 
limit for the kinetic temperature
of the gas. The excited rotational levels, $J=2,3$ and $4$,
fit on a line with $T_{24}=885\pm270$ K, possibly 
indicating processes like UV pumping, H$_2$ formation pumping, and
shock excitation.
\\
\\

Fig. 8. FUSE continuum-normalized H$_2$ absorption profiles 
toward 9 background sources are shown.
For each sight line, an H$_2$ absorption profile from
the Lyman band (labeled in the lower right corner) is
plotted and compared to the H\,{\sc i} 21cm emission 
(labeled in the upper right corner) from either
Green Bank data (GB), Effelsberg data (EB), or 
Leiden-Dwingeloo data (LD; see Wakker et al.\,2001;
Murphy et al.\,1996; Hartmann \& Burton 1997). The two major absorption 
components along each sight line from local Galactic 
gas near zero velocities and from halo gas at intermediate velocities
between $25 \leq |v_{\rm LSR}| \leq 100$ km\,s$^{-1}$ are marked with
dashed lines. IVC H$_2$ absorption that matches 
the H\,{\sc i} 21cm emission line data is present
in each of the absorption profiles.
Stellar background sources are labeled with a star
symbol behind the object name. Lines from other
transitions or 
species are marked with `b'. The abbreviation `N.F.'
is used for the normalized flux.   
\\
\\

Fig. 9. Composite H$_2$ absorption profiles for 
NGC\,7714, Mrk\,357, and Mrk\,618,
representing examples of tentative H$_2$ 
detections at IVC velocities in FUSE 
data with low S/N. 
At least three H$_2$ lines per sight line have been stacked on top
of each other in order to study the velocity extent of H$_2$ in these
directions. 
As in the previous figure, H\,{\sc i} 21cm data
is plotted above each H$_2$ absorption profile. The          
local Galactic components and the IVC components are marked with dashed lines.
In the three examples shown, H$_2$ appears to be present
at IVC velocities, but more accurate absorption line data are required 
to confirm these detections.
\\
\\

Fig. 10. North- and south-polar sky distribution of the 56 sight lines that sample H\,{\sc i} IVC
gas for $|b| \geq 25$ and $25 \leq |v_{\rm LSR}| \leq 100$
km\,s$^{-1}$, plotted on top of the H\,{\sc i} 21cm IVC sky
from the Leiden-Dwingeloo Survey
(Hartmann \& Burton 1997; H\,{\sc i} data plotted for 
$-30\leq v_{\rm LSR} \leq -90$ km\,s$^{-1}$; see 
also Fig.\,1 for details).
Definitive IVC H$_2$ detections are marked with filled boxes,
while tentative detections are labeled with filled triangles;
non-detections are marked with open circles (see also Table 2).
The H$_2$ absorption generally follows the H\,{\sc i} emission for
regions with high H\,{\sc i} column densities.     
Note that the H\,{\sc i} data show IVC gas 
solely at negative radial velocities,
while for some lines of sight that are labeled here the main IVC
component is at {\it positive} velocities (see Table\,2).
The position of the Magellanic Clouds (LMC+SMC) is also indicated.
\\
\\

Fig. 11. The logarithmic H$_2$ fraction, log $f=$\,log $[2N($H$_2)/(N$(H\,{\sc i}$)+2N($H$_2))]$,
is plotted against the logarithmic H\,{\sc i} 21cm column density, log $N($H\,{\sc i}$)$, 
for the 61 IVC components listed in Table 2.
H$_2$ detections are plotted as filled
circles, possible detections and the non-detections 
have open symbols (open triangles for the possible detections, open
circles for the non-detections; note that these symbols
represent upper
limits for log $f$). All IVC components
in which H$_2$ is clearly detected (14 components) 
have log $f\leq-3$.

\clearpage
\newpage
\begin{deluxetable}{lrcl}
\tabletypesize{\normalsize}
\tablecaption{H$_2$ Equivalent Widths$^{\rm a}$ for the IV Arch towards PG\,1351+640
\label{tbl-2}}
\tablewidth{0pt}
\tablehead{
\colhead{Line} & \colhead{$\lambda_{\rm vac}$\,$^{\rm b}$} & \colhead{log\,$\lambda f^{\rm b}$}
& \colhead{$W_{\lambda}$\,$_{\rm IVC}$} \\
\colhead{} & \colhead{[\AA]}   & \colhead{}   & \colhead{[m\AA]} 
}
\startdata
\hline
\multicolumn{4}{c}{$J=0$; log $N(0)=15.80\pm0.21$}\\
\hline
R(0),8-0         &   1001.828  &     1.432 &   $37\pm12$ \\
R(0),0-0$^{\rm c,d}$ &   1008.559  &     1.647 &   $98\pm23$ \\
R(0),6-0         &   1024.376  &     1.473 &   $54\pm11$ \\
R(0),4-0         &   1049.367  &     1.383 &   $35\pm9$ \\
R(0),2-0         &   1077.142  &     1.111 &   $32\pm9$ \\
R(0),1-0         &   1092.200  &     0.802 &   $33\pm7$ \\
R(0),0-0         &   1108.130  &     0.275 &   $28\pm13$ \\
\hline
\multicolumn{4}{c}{$J=1$; log $N(1)=16.23\pm0.14$}\\
\hline
P(1),8-0         & 1003.302    &     0.948 &   $57\pm9$ \\
R(1),5-0         & 1037.150    &     1.271 &   $83\pm12$ \\
R(1),4-0         & 1049.960    &     1.225 &   $53\pm20$ \\ 
P(1),4-0         & 1051.033    &     0.902 &   $50\pm9$  \\ 
P(1),3-0         & 1064.606    &     0.805 &   $47\pm11$ \\
R(1),2-0         & 1077.702    &     0.919 &   $51\pm9$  \\
P(1),2-0         & 1078.929    &     0.624 &   $54\pm11$ \\
R(1),1-0         & 1092.737    &     0.618 &   $49\pm8$ \\
P(1),1-0         & 1094.057    &     0.340 &   $57\pm9$ \\
R(1),0-0         & 1108.636    &     0.086 &   $48\pm9$ \\
\hline
\multicolumn{4}{c}{$J=2$; log $N(2)=14.91\pm0.33$}\\
\hline
R(2),8-0         & 1003.989    &     1.232 &   $\leq30$ \\
P(2),8-0         & 1005.398    &     0.998 &   $\leq27$ \\
R(2),0-0$^{\rm c}$ & 1009.023   &     1.208 &   $33\pm9$ \\
Q(2),0-0$^{\rm c}$ & 1010.938   &     1.385 &   $43\pm12$ \\
R(2),7-0$^{\rm e}$ & 1014.980   &     1.285 &   $30\pm10$ \\
P(2),7-0         & 1016.466    &     1.007 &   $36\pm8$ \\
R(2),4-0         & 1051.498    &     1.168 &   $29\pm11$ \\
P(2),4-0         & 1053.284    &     0.982 &   $\leq68$ \\
R(2),3-0         & 1064.994    &     1.069 &   $38\pm9$ \\
P(2),1-0         & 1096.444    &     0.420 &   $\leq57$ \\
\hline
\multicolumn{4}{c}{$J=3$; log $N(3)=15.33\pm0.26$}\\
\hline
P(3),7-0         & 1019.507   &      1.050  &  $42\pm10$ \\
P(3),6-0         & 1031.195   &      1.055  &  $54\pm9$ \\
P(3),5-0         & 1043.504   &      1.060  &  $56\pm10$ \\
R(3),4-0         & 1053.975   &      1.137  &  $\leq58$ \\
P(3),4-0         & 1056.471   &      1.006  &  $47\pm9$ \\
R(3),3-0         & 1067.478   &      1.066  &  $41\pm6$ \\
P(3),3-0         & 1070.141   &      0.910  &  $\leq35$ \\
P(3),1-0         & 1099.792   &      0.439  &  $29\pm11$ \\
\hline
\multicolumn{4}{c}{$J=4$; log $N(4)\leq14.64$}\\
\hline
P(4),6-0    &  1035.184  &     1.056 &    $\leq33$ \\
R(4),5-0    &  1044.543  &     1.195 &    $\leq44$ \\
R(4),3-0    &  1070.899  &     1.012 &    $\leq27$ \\
\hline
\multicolumn{4}{c}{log $N$(Total)$=16.43\pm0.14$}\\
\hline
\enddata
\tablenotetext{a}{Equivalent widths and $1 \sigma$ errors (or $3 \sigma$ upper limits) 
are listed for the $-50$ km\,s$^{-1}$ IVC absorption.}
\tablenotetext{b}{Vacuum wavelengths and oscillator strengths from Abgrall \& Roueff (1989).}
\tablenotetext{c}{Line from the Werner band.}
\tablenotetext{d}{Blended together with Werner Q(1),0-0 \& Lyman P(3),8-0.}
\tablenotetext{e}{Blended together with Werner Q(4),0-0.}
\end{deluxetable}

\clearpage
\newpage
\begin{deluxetable}{llccclccrrl}
\rotate
\tabletypesize{\scriptsize}
\tablecaption{FUSE Survey of H$_2$ in Intermediate-Velocity Clouds
\label{tbl-2}}
\tablewidth{0pt}
\tablehead{
\colhead{Object} & \colhead{Type} & \colhead{$l$} & \colhead{$b$} & \colhead{$v_{\rm IVC}$\,$^{\rm a}$} 
& \colhead{IVC Name}
& \colhead{log $N$(H\,{\sc i})$_{\rm IVC}$} & \colhead{IVC H$_2$ Status} & 
\colhead{log $N($H$_2)_{\rm IVC}$} & \colhead{log $f^{\rm f}$\,$_{\rm IVC}$} & \colhead{Remarks}\\
& & & & \colhead{[km\,s$^{-1}$]} 
}
\startdata
Mrk\,509     & QSO     &  35.97 & $-$29.86 & $+60$ & Complex gp & 19.51$^{\rm b}$ & yes                & $14.92$     & $-
4.3$     & \nodata\\
vZ\,1128     & Star    &  42.50 & $+$78.68 & $+29$ & \nodata    & 19.31$^{\rm d}$ & no evidence        & $\leq15.04$ & 
$\leq-4.0$ & \nodata\\
Mrk\,1513    & Seyfert &  63.67 & $-$29.07 & $-29$ & \nodata    & 19.37$^{\rm b}$ & no evidence        & $\leq16.43$ & 
$\leq-2.6$ & \nodata\\
Mrk\,501     & BLLac   &  63.60 & $+$38.86 & $-38$ & \nodata    & 19.17$^{\rm c}$ & no evidence        & $\leq17.08$ & 
$\leq-1.8$ & low S/N$^{\rm m}$\\
PG\,1444+407 & QSO     &  69.90 & $+$62.72 & $-30$ & \nodata    & 19.53$^{\rm b}$ & no evidence        & $\leq17.38$ 
& $\leq-1.9$ & low S/N\\
Mrk\,487     & Galaxy  &  87.84 & $+$49.03 & $-85$ & IV Arch (IV\,15)  & 19.16$^{\rm b}$ & no evidence & $\leq17.43$ 
& $\leq-1.4$ & low S/N\\
NGC\,7714    & Seyfert &  88.22 & $-$55.56 & $-50$ & \nodata    & 19.40$^{\rm b}$ & possibly           & $\leq17.03$ & 
$\leq-2.1$ & low S/N\\
Mrk\,876     & QSO     &  98.27 & $+$40.38 & $-30$ & Draco      & 19.84$^{\rm c}$ & yes                & $15.57$     & $-4.0$     
& \nodata\\
NGC\,7673    & Galaxy  &  99.25 & $-$35.40 & $-53$ & PP Arch    & 19.25$^{\rm b}$ & no evidence        & $\leq17.14$ & 
$\leq-1.8$ & low S/N\\
Mrk\,817     & Seyfert & 100.30 & $+$53.48 & $-40$ & IV Arch    & 19.33$^{\rm c}$ & no evidence        & $\leq14.95$ & 
$\leq-4.1$ & \nodata\\
NGC\,5461    & M\,101-H\,{\sc ii} & 101.89 & $+$59.76 & $-51$ & IV Arch (IV\,19) & 19.49$^{\rm d}$ & no evidence & 
$\leq16.32$ & $\leq-2.9$ & low S/N\\
Mrk\,335     & Seyfert & 108.76 & $-$41.42 & $-27$ & \nodata    & 20.00$^{\rm b}$ & possibly           & $\leq15.82$ & 
$\leq-3.9$ & low S/N\\
Mrk\,59      & Galaxy  & 111.54 & $+$82.12 & $-44$ & IV Arch    & 19.31$^{\rm d}$ & yes                & $14.72$     & $-4.3$     
& \nodata\\
PG\,1351+640 & QSO     & 111.89 & $+$52.02 & $-47$ & IV Arch (IV\,19) & 20.05$^{\rm c}$ & yes          & $16.43$     & 
$-3.3$     & see \S3\\
             &         &        &          & $-74$ & IV Arch (IV\,9)  & 19.20$^{\rm c}$ & no evidence  & $\leq15.18$ & $\leq-3.7$ & 
\nodata\\
HD\,121800   & Star    & 113.01 & $+$49.76 & $-70$ & IV Arch   & 19.86$^{\rm c}$ & yes                 & $14.29$     & $-
5.3$     & \nodata\\
Mrk\,279     & Seyfert & 115.04 & $+$46.86 & $-40$ & LLIV Arch  & 19.80$^{\rm c}$ & no evidence        & $\leq14.85$ 
& $\leq-4.6$ & \nodata\\
             &         &        &          & $-76$ & IV Arch (IV\,9)  & 19.42$^{\rm c}$ & no evidence  & $\leq14.85$ & $\leq-4.3$ & 
\nodata\\
PG\,1259+593 & QSO     & 120.56 & $+$58.05 & $-54$ & IV Arch    & 19.50$^{\rm c}$ & yes                & $14.10^{\rm g}$ 
& $-5.1$  & \nodata\\
PG\,0052+251 & QSO     & 123.91 & $-$37.44 & $-39$ & \nodata    & 19.82$^{\rm b}$ & possibly           & $\leq17.12$ & 
$\leq-2.4$ & low S/N\\
Mrk\,352     & Seyfert & 125.03 & $-31.01$ & $-28$ & \nodata    & 20.33$^{\rm b}$ & possibly           & $\leq17.05$ & 
$\leq-3.0$ & low S/N\\
Mrk\,205     & Seyfert & 125.45 & $+$41.67 & $-48$ & LLIV Arch  & 19.71$^{\rm b}$ & possibly           & $\leq16.69$ & 
$\leq-2.7$ & low S/N\\
3C\,249.1 & QSO     & 130.39 & $+$38.55 & $-50$ & LLIV Arch    & 19.80$^{\rm b}$ &    possibly           &             $\leq 
17.00$ & $\leq-2.5$            & low S/N\\
Mrk\,357     & Seyfert & 132.20 & $-$39.14 & $-41$ & \nodata    & 19.58$^{\rm d}$ & possibly           & $\leq17.16$ & 
$\leq-2.1$ & low S/N\\
NGC\,3516    & Seyfert & 133.24 & $+$42.40 & $-47$ & LLIV Arch  & 19.91$^{\rm b}$ & possibly           & $\leq17.23$ 
& $\leq-2.4$ & low S/N\\
Mrk\,209     & Galaxy  & 134.15 & $+$68.08 & $-55$ & IV Arch    & 19.61$^{\rm b}$ & no evidence        & $\leq16.79$ & 
$\leq-2.5$ & \nodata\\
             &         &        &          & $-100$ & IV Arch (IV\,4) & 19.06$^{\rm b}$ & no evidence  & $\leq16.79$ & $\leq-2.0$ 
& \nodata\\
PG\,0804+761 & QSO     & 138.28 & $+$31.03 & $-58$ & LLIV Arch  & 19.56$^{\rm c}$ & yes                & $14.71^{\rm 
h}$ & $-4.5$ & \nodata\\
NGC\,3690    & Galaxy  & 141.91 & $+$55.41 & $-54$ & IV Arch    & 19.56$^{\rm b}$ & no evidence        & $\leq16.77$ 
& $\leq-2.5$ & \nodata\\
HS\,0624+6907 & QSO    & 145.71 & $+$23.35 & $-27$ & LLIV Arch  & 20.15$^{\rm c}$ & no evidence        & 
$\leq17.41$ & $\leq-2.4$ & low S/N\\
PG\,0832+675 & Star    & 147.75 & $+$35.01 & $-50$ & LLIV Arch  & 19.99$^{\rm c}$ & yes                & $15.76$     & $-
3.9$     & \nodata\\
NGC\,4151    & Seyfert & 155.08 & $+$75.06 & $-29$ & IV Arch (IV\,26) & 20.20$^{\rm b}$ & yes          & $15.36$     & 
$-4.5$     & \nodata\\
NGC\,3310    & Seyfert & 156.60 & $+$54.06 & $-47$ & IV Arch    & 19.80$^{\rm b}$ & yes                & $14.95$     & $-
4.5$     & \nodata\\
Mrk\,9       & Seyfert & 158.36 & $+$28.75 & $-40$ & LLIV Arch  & 19.60$^{\rm b}$ & possibly           & $\leq16.48$ & 
$\leq-2.8$ & strong local H$_2$\\
NGC\,4214    & Galaxy  & 160.24 & $+$78.07 & $-33$ & IV Arch (IV\,26) & 20.13$^{\rm d}$ & possibly     & $\leq15.74$ 
& $\leq-4.1$ & \nodata\\
Mrk\,116     & Galaxy  & 160.53 & $+$44.84 & $-39$ & LLIV Arch  & 19.48$^{\rm c}$ & possibly           & $\leq16.35$ & 
$\leq-2.8$ & low S/N\\
Mrk\,106     & Seyfert & 161.14 & $+$42.88 & $-40$ & LLIV Arch  & 19.35$^{\rm c}$ & possibly           & $\leq17.48$ & 
$\leq-1.6$ & low S/N\\
NGC\,1068    & Seyfert & 172.10 & $-$51.93 & $-58$ & \nodata    & 19.08$^{\rm b}$ & no evidence        &             
$\leq15.26$ & $\leq-3.5$         & \nodata\\
PG\,0953+414 & QSO     & 179.79 & $+$51.71 & $-49$ & IV Arch (IV\,16) & 19.33$^{\rm c}$ & no evidence  & 
$\leq15.07$ & $\leq-4.0$ & \nodata\\
Mrk\,421     & BLLac   & 179.83 & $+$65.03 & $-61$ & IV Arch (IV\,26) & 19.78$^{\rm b}$ & no evidence  & 
$\leq14.93$ & $\leq-4.5$ & \nodata\\
PG\,0947+396 & QSO     & 182.85 & $+$50.75 & $-66$ & IV Arch (IV\,16) & 19.68$^{\rm b}$ & no evidence  &             
$\leq17.11$ &  $\leq-2.3$       & low S/N\\
HD\,93521    & Star    & 183.14 & $+$62.15 & $-62$ & IV Arch    & 19.58$^{\rm e}$ & yes                & $14.60^{\rm i}$ & 
$-4.7$ & ORFEUS \\
NGC\,3991    & Galaxy  & 185.68 & $+$77.20 & $-53$ & IV Arch (IV\,18) & 19.86$^{\rm e}$ & no evidence  & 
$\leq17.26$ & $\leq-2.3$ & low S/N\\
Ton\,1187    & QSO     & 188.33 & $+$55.38 & $-27$ & IV Arch    & 19.55$^{\rm b}$ & no evidence        & $\leq17.40$ 
& $\leq-1.9$ & low S/N\\
             &         &        &          & $-69$ & IV Arch    & 19.47$^{\rm b}$ & no evidence        & $\leq17.40$ & $\leq-1.8$ & 
low S/N\\
PG\,1001+291 & QSO     & 200.09 & $+$53.20 & $-32$ & IV Arch (IV\,18) & 20.12$^{\rm b}$ & possibly     &             
$\leq17.18$ & $\leq-2.6$      & low S/N\\
HE\,0238-1904 & QSO    & 200.48 & $-$63.63 & $-42$ & \nodata    & 19.33$^{\rm d}$ & no evidence        & $\leq17.10$ 
& $\leq-1.9$ & low S/N\\
Mrk\,36      & Galaxy  & 201.76 & $+$66.49 & $-55$ & IV Arch (IV\,18)  & 19.83$^{\rm d}$ & possibly    & $\leq17.15$ & 
$\leq-2.4$ & low S/N\\
NGC\,3504    & Galaxy  & 204.60 & $+$66.04 & $-52$ & IV Arch (IV\,18)  & 19.85$^{\rm d}$ &  possibly   & $\leq16.33$ 
& $\leq-3.2$ & low S/N\\
Mrk\,618     & Seyfert & 206.72 & $-$34.66 & $-27$ & \nodata    & 19.56$^{\rm b}$ &  possibly          & $\leq17.15$ & 
$\leq-2.1$ & low S/N\\
PG\,1116+215 & QSO     & 223.36 & $+$68.21 & $-42$ & IV Spur    & 19.83$^{\rm c}$ & yes                &             $15.27$ & 
$-4.3$        & \nodata\\
Mrk\,734     & Seyfert & 244.75 & $+$63.94 & $-42$ & IV Spur    & 19.81$^{\rm b}$ & no evidence        & $\leq16.90$ & 
$\leq-2.6$ & low S/N\\
PKS\,0558-504 & QSO    & 257.96 & $-$28.57 & $+26$ &  \nodata   & 20.31$^{\rm e}$ & [no evidence]      & \nodata     
& \nodata    & no atomic absorption seen\\
              &        &        &          & $+71$ & \nodata    & 19.62$^{\rm e}$ & [no evidence]      & \nodata     & \nodata    & 
no atomic absorption seen\\
HD\,100340   & Star    & 258.85 & $+$61.23 & $-29$ & IV Spur    & 19.98$^{\rm d}$ & yes                & $15.98$     & $-3.7$     
& MWRS$^{\rm k}$\\
NGC\,1705    & Galaxy  & 261.08 & $-$38.74 & $+87$ & \nodata    & 19.64$^{\rm e}$ & [no evidence]      & \nodata     & 
\nodata    & no atomic absorption seen\\
PG\,1211+143 & QSO     & 267.55 & $+$74.32 & $-35$ & IV Spur    & 19.94$^{\rm b}$ & no evidence        & $\leq15.23$ 
& $\leq-4.4$ & \nodata\\
Mrk\,771     & Seyfert & 269.44 & $+$81.74 & $-29$ & IV Spur    & 19.93$^{\rm b}$ & no evidence        & $\leq16.78$ & 
$\leq-2.8$ & low S/N\\
NGC\,3783    & Seyfert & 287.46 & $+$22.95 & $+34$ & \nodata    & 19.44$^{\rm b}$ & possibly           & $\leq16.52$ & 
$\leq-2.6$ & \nodata \\
             &         &        &          & $+62$ & \nodata    & 20.00$^{\rm b}$ & no evidence        & $\leq15.84^{\rm k}$ & $\leq-
3.9$ & \nodata \\
3C\,273      & QSO     & 289.95 & $+$64.36 & $+25$ & \nodata    & 19.42$^{\rm b}$ & yes                & $15.71^{\rm l}$     
& $-3.4$ & \nodata \\
HE\,1228+0131 & QSO    & 291.26 & $+$63.66 & $+27$ & \nodata    & 19.26$^{\rm c}$ & no evidence        & 
$\leq15.96$ & $\leq-3.0$ & low S/N \\
Tol\,1247-232 & Galaxy & 302.60 & $+$39.30 & $+54$ & \nodata    & 19.40$^{\rm d}$ & no evidence        & 
$\leq17.24$ & $\leq-1.9$ & low S/N\\
ESO\,141-G55 & Seyfert & 338.18 & $-$26.71 & $-45$ & \nodata    & 19.27$^{\rm e}$ & no evidence        & $\leq17.02$ 
& $\leq-2.0$ & \nodata \\

\enddata
\tablenotetext{a}{LSR velocities, based on 21cm data, are listed; $^{\rm b}$H\,{\sc i} 
21cm data from the Greenbank 140ft telescope (Murphy et al.\,1996); $^{\rm c}$ H\,{\sc i} 
21cm data from the Effelsberg 100m telescope (Wakker et al.\,2001); $^{\rm d}$
H\,{\sc i} 21cm data from the Leiden-Dwingeloo Survey (Hartmann \& Burton 1997);
$^{\rm e}$H\,{\sc i} 21cm data from the Villa Elisa telescope (Arnal et al.\,2000);
$^{\rm f }$log $f=$ log $[2N($H$_2)/(N$(H\,{\sc i}$)+2N($H$_2))]$;
$^{\rm g}$Richter et al.\,(2001c); $^{\rm h}$Richter et al.\,(2001a);
$^{\rm i}$Gringel et al.\,(2000); $^{\rm j}$Observed through the medium-resolution aperture;
$^{\rm k}$See also Sembach et al.\,(2001a); $^{\rm l}$See also Sembach et al.\,(2001b) and \S4;
$^{\rm m}$FUSE spectra with an average S/N$<9$ (per resolution element); 
see also Wakker et al.\,(2002).}
\end{deluxetable}

\clearpage
\newpage
\begin{deluxetable}{lrcl}
\tabletypesize{\normalsize}
\tablecaption{Selected H$_2$ Equivalent Widths$^{\rm a}$ for IVCs
\label{tbl-2}}
\tablewidth{0pt}
\tablehead{
\colhead{Line} & \colhead{$\lambda_{\rm vac}$\,$^{\rm b}$} & \colhead{log\,$\lambda f^{\rm b}$}
& \colhead{$W_{\lambda}$\,$_{\rm IVC}$} \\
\colhead{} & \colhead{[\AA]}   & \colhead{}   & \colhead{[m\AA]}
}
\startdata
\hline
\multicolumn{4}{c}{Mrk\,509 - Complex gp}\\
\hline
R(0),0-0$^{\rm c,d}$        &   1008.559  &     1.647 &   $29\pm5$ \\
R(1),4-0                    &   1049.960  &     1.225 &   $36\pm6$ \\
R(2),7-0                    &   1014.980  &     1.285 &   $26\pm8$ \\
R(3),6-0                    &   1028.988  &     1.243 &   $26\pm4$ \\
R(3),4-0                    &   1053.975  &     1.137 &   $18\pm5$ \\
P(3),4-0                    &   1056.471  &     1.006 &   $20\pm4$ \\
\hline
\multicolumn{4}{c}{Mrk\,876 - Draco}\\
\hline
R(2),8-0                    &   1003.989  &     1.232 &   $26\pm7$ \\
Q(2),0-0$^{\rm c}$          &   1010.938  &     1.385 &   $31\pm6$ \\
R(2),3-0                    &   1064.994  &     1.069 &   $30\pm6$ \\
R(3),0-0$^{\rm c}$          &   1010.128  &     1.151 &   $25\pm5$ \\
P(3),5-0                    &   1043.504  &     1.060 &   $24\pm6$ \\
R(3),4-0                    &   1053.975  &     1.137 &   $20\pm5$ \\
\hline
\multicolumn{4}{c}{Mrk\,59 - IV Arch}\\
\hline
R(1),8-0                    &   1002.457  &     1.256 &   $35\pm7$ \\
R(1),4-0                    &   1049.960  &     1.225 &   $38\pm9$ \\
P(1),4-0                    &   1051.033  &     0.902 &   $20\pm6$ \\
R(1),2-0                    &   1077.702  &     0.919 &   $23\pm8$ \\
P(1),2-0                    &   1078.929  &     0.624 &   $21\pm6$ \\
P(3),5-0                    &   1043.504  &     1.060 &   $23\pm9$ \\
\hline
\multicolumn{4}{c}{HD\,121800 - IV Arch}\\
\hline
R(1),8-0                    &   1002.457  &     1.256 &   $14\pm4$ \\
R(1),7-0                    &   1013.441  &     1.307 &   $17\pm4$ \\
R(1),4-0                    &   1049.960  &     1.225 &   $25\pm6$ \\
P(1),4-0                    &   1051.033  &     0.902 &   $12\pm4$ \\
P(1),3-0                    &   1064.606  &     0.805 &   $18\pm4$ \\
R(1),2-0                    &   1077.702  &     0.919 &   $11\pm3$ \\
\hline
\multicolumn{4}{c}{PG\,1259+593 - IV Arch$^{\rm e}$}\\
\hline
R(1),8-0                    &   1002.457  &     1.256 &   $21\pm5$ \\
Q(1),0-0$^{\rm c}$          &   1009.770  &     1.384 &   $19\pm4$ \\
P(1),7-0                    &   1014.332  &     0.960 &   $13\pm4$ \\
R(1),4-0                    &   1049.960  &     1.225 &   $14\pm4$ \\
P(1),4-0                    &   1051.033  &     0.902 &   $10\pm3$ \\
Q(2),0-0$^{\rm c}$          &   1010.938  &     1.385 &   $14\pm4$ \\
\hline
\multicolumn{4}{c}{PG\,0832+675 - LLIV Arch}\\
\hline
P(2),6-0                    &   1028.108  &     1.053 &   $31\pm9$ \\
R(2),5-0                    &   1038.690  &     1.221 &   $25\pm7$ \\
R(2),4-0                    &   1051.498  &     1.168 &   $45\pm12$ \\
P(3),5-0                    &   1043.504  &     1.060 &   $42\pm7$ \\
R(3),4-0                    &   1053.975  &     1.137 &   $42\pm8$ \\
P(3),4-0                    &   1056.471  &     1.006 &   $30\pm8$ \\
\hline
\multicolumn{4}{c}{NGC\,4151 - IV Arch (IV\,26)} \\
\hline
P(1),5-0                    &   1038.158  &     0.956 &   $64\pm12$ \\
P(1),4-0                    &   1051.033  &     0.902 &   $73\pm16$ \\
P(3),6-0                    &   1031.195  &     1.055 &   $43\pm5$ \\
R(3),5-0                    &   1041.159  &     1.222 &   $44\pm5$ \\
P(3),5-0                    &   1043.504  &     1.060 &   $55\pm8$ \\
R(4),5-0                    &   1044.543  &     1.195 &   $21\pm7$ \\
\hline
\multicolumn{4}{c}{NGC\,3310 - IV Arch}\\
\hline
P(2),8-0                    &   1005.398  &     0.998 &   $33\pm8$ \\
Q(2),0-0$^{\rm c}$          &   1010.938  &     1.385 &   $50\pm9$ \\
P(2),4-0                    &   1053.284  &     0.982 &   $45\pm9$ \\
R(3),7-0                    &   1017.427  &     1.263 &   $42\pm8$ \\
R(3),4-0                    &   1053.975  &     1.137 &   $26\pm6$ \\
P(3),4-0                    &   1056.471  &     1.006 &   $24\pm5$ \\
\hline
\multicolumn{4}{c}{PG\,1116+215 - IV Spur}\\
\hline
R(0),4-0                    &   1049.367  &     1.383 &   $46\pm5$ \\
R(0),2-0                    &   1077.142  &     1.111  &  $41\pm5$ \\  
R(1),4-0                    &   1049.960  &     1.225 &   $68\pm8$ \\
R(1),2-0                    &   1077.702  &     0.919 &   $47\pm6$ \\
P(1),2-0                    &   1078.929  &     0.624 &   $41\pm9$ \\
Q(2),0-0$^{\rm c}$          &   1010.938  &     1.385 &   $38\pm5$ \\
\hline
\multicolumn{4}{c}{HD\,100340 - IV Spur}\\
\hline
P(1),5-0                    &   1038.158  &     0.956 &   $74\pm13$ \\
P(1),2-0                    &   1078.929  &     0.624 &   $64\pm7$ \\
R(2),4-0                    &   1051.498  &     1.168 &   $20\pm4$ \\
R(2),3-0                    &   1064.994  &     1.069 &   $27\pm5$ \\
R(3),4-0                    &   1053.975  &     1.137 &   $17\pm5$ \\
P(3),5-0                    &   1043.504  &     1.060 &   $9\pm3$ \\
\hline
\enddata
\tablenotetext{a}{Six equivalent widths per sight line and $1 \sigma$ errors (or $3 \sigma$ upper limits) are listed.}
\tablenotetext{b}{Vacuum wavelengths and oscillator strengths from Abgrall \& Roueff (1989).}
\tablenotetext{c}{Line from the Werner band.}
\tablenotetext{d}{Blended together with Werner Q(1),0-0 \& Lyman P(3),8-0.}
\tablenotetext{e}{See also Richter et al.\,(2001c).}
\end{deluxetable}

\clearpage
\newpage
\begin{deluxetable}{llrrrrrcl}
\tabletypesize{\scriptsize}
\tablecaption{Newly Measured H$_2$ Column Densities in IVCs 
\label{tbl-2}}
\tablewidth{0pt}
\tablehead{
\colhead{Object} & \colhead{IVC Name} & \colhead{log $N(0)$} & \colhead{log $N(1)$}
& \colhead{log $N(2)$} & \colhead{log $N(3)$} & \colhead{log $N(4)$} & 
\colhead{$b$} & \colhead{Total log $N$(H$_2$)$^{\rm b}$}\\
\colhead{} & \colhead{} & \colhead{} & \colhead{} & \colhead{} &
\colhead{} & \colhead{} & \colhead{[km\,s$^{-1}$]} & \colhead{}
}
\startdata
Mrk\,509 & Complex gp        & $\leq 13.00$ & $14.60$ & $14.33$      & $14.35$ & $\leq14.09$ & $5.5$ & $14.92\pm0.46$ 
\\
Mrk\,876 & Draco             & $14.93$      & $15.38$ & $14.39$      & $14.32$ & $\leq14.35$ & $5.4$ & $15.57\pm0.23$ \\
Mrk\,59   & IV Arch          & $\leq 14.22$ & $14.72$ & $\leq 14.10$ & $\leq 14.55$ & $\leq 14.36$ & $4.1$ & 
$14.72\pm0.28$ \\
HD\,121800 & IV Arch         & $\leq 13.94$ & $14.29$ & $\leq 14.27$ & $\leq 14.23$ & $\leq 14.19$ & $3.0$ & 
$14.29\pm0.62$ \\
PG\,1259+593 & IV Arch       & $\leq 13.69$ & $14.10$ & $\leq 13.85$ & $\leq 13.87$ & $\leq 13.96$ & L$^{\rm a}$ & 
$14.10\pm0.19$ \\
PG\,0832+675 & LLIV Arch     & $15.47$ & $15.25$ & $14.57$ & $14.80$ & $\leq 14.30$ & $5.3$ & $15.76\pm0.31$ \\
NGC\,4151 & IV Arch (IV\,26) & $14.86$ & $14.98$ & $14.39$ & $14.58$ & $\leq 14.14$ & L$^{\rm a}$ & 
$15.36\pm0.11$ \\
NGC\,3310 & IV Arch          & $\leq 16.50$ & $\leq 16.11$ & $14.76$ & $14.49$ & $\leq 14.35$ & $7.4$ & $14.95\pm0.77$ 
\\
PG\,1116+215 & IV Spur       & $14.59$      & $15.02$      & $14.45$ & $14.18$ & $\leq 14.24$ & $7.9$ & $15.27\pm0.34$ \\ 
HD\,100340     & IV Spur     & $15.66$ & $15.67$ & $14.34$ & $13.99$ & $\leq 13.85$ & $6.6$ & $15.98\pm0.83$ \\
\enddata
\tablenotetext{a}{Derived column density assumes that the data points fall on the linear part of the curve of growth.}
\tablenotetext{b}{$1\sigma$ error is given.}
\end{deluxetable}

\clearpage
\newpage
\includegraphics{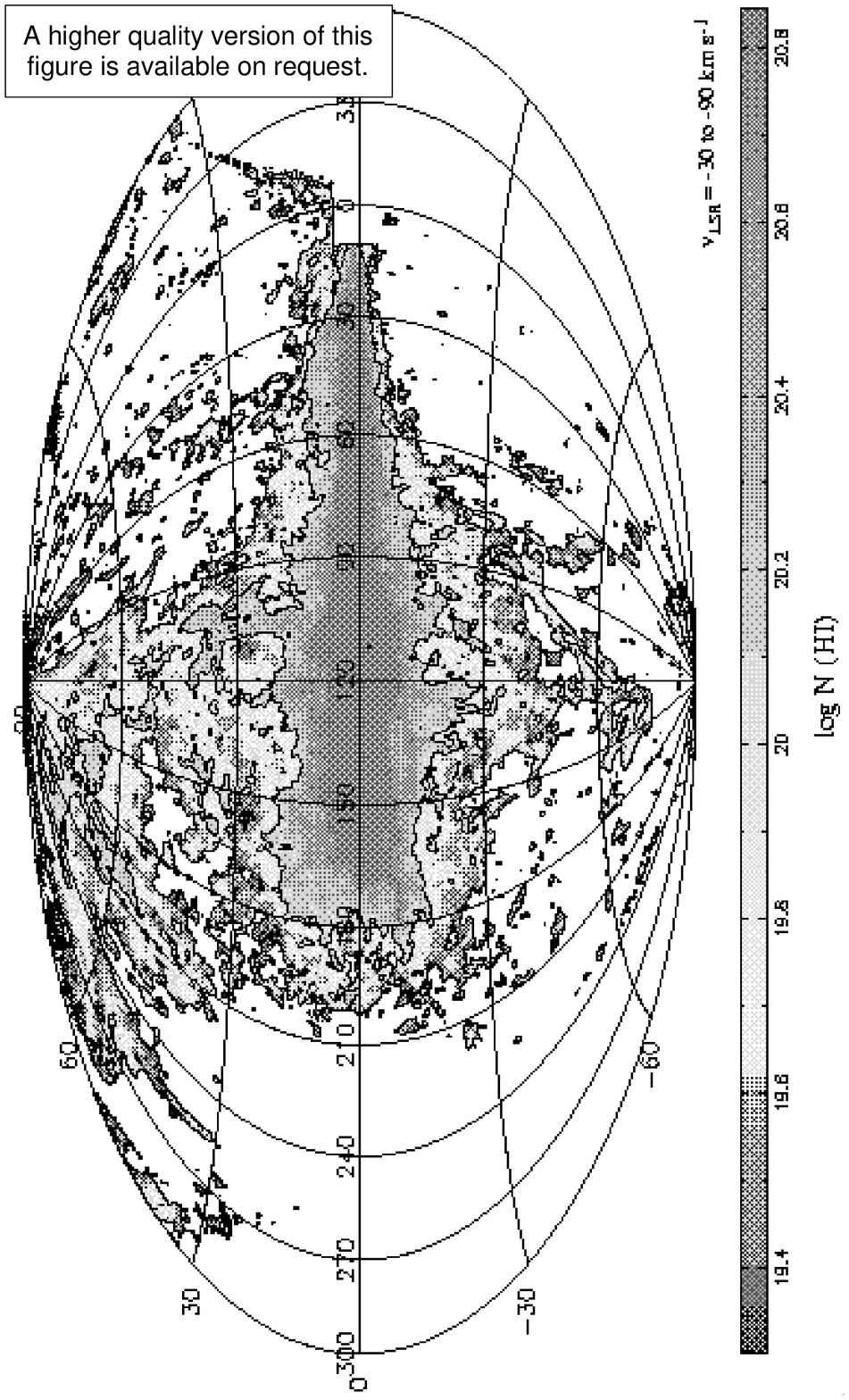}
\figcaption[]{}

\clearpage
\newpage
\includegraphics{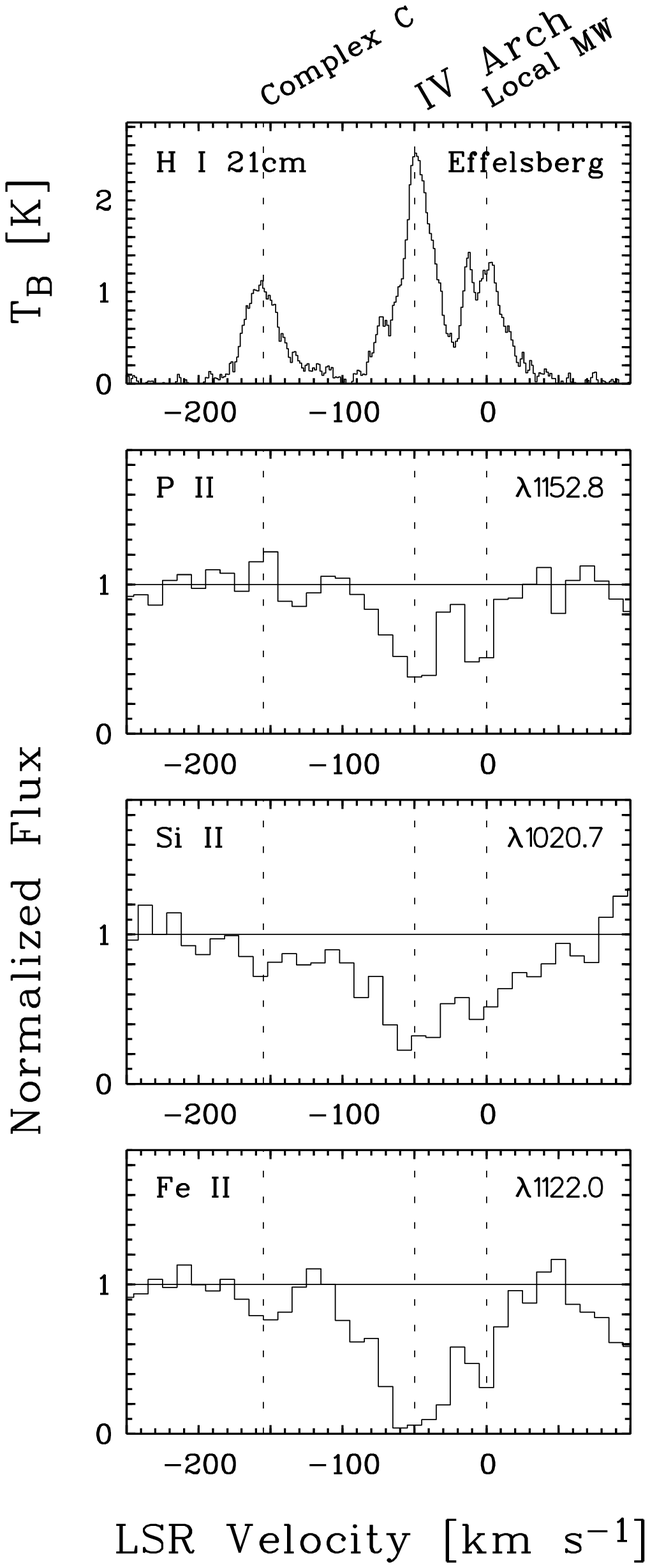}
\figcaption[]{}

\clearpage
\newpage
\includegraphics{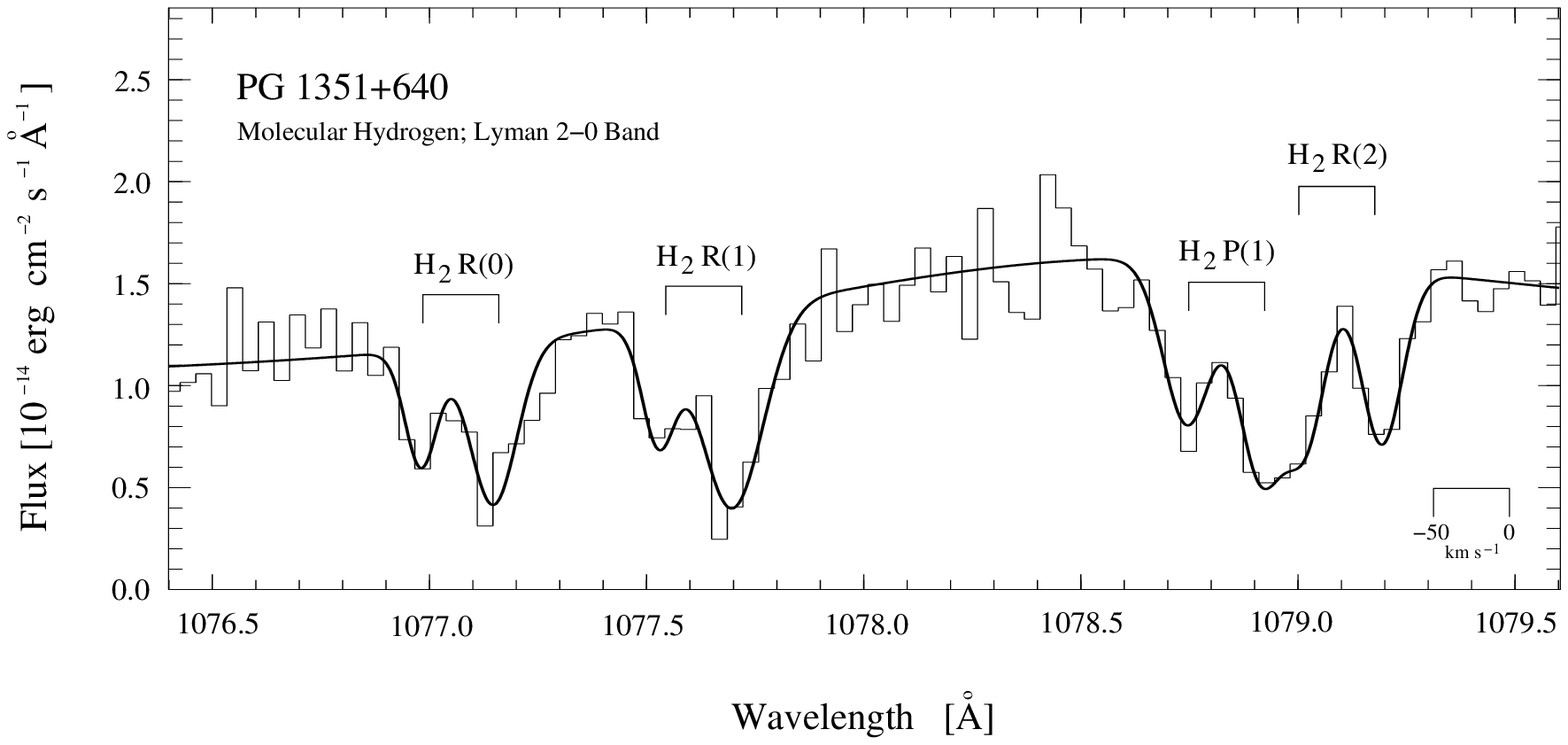}
\figcaption[]{}

\clearpage
\newpage
\includegraphics{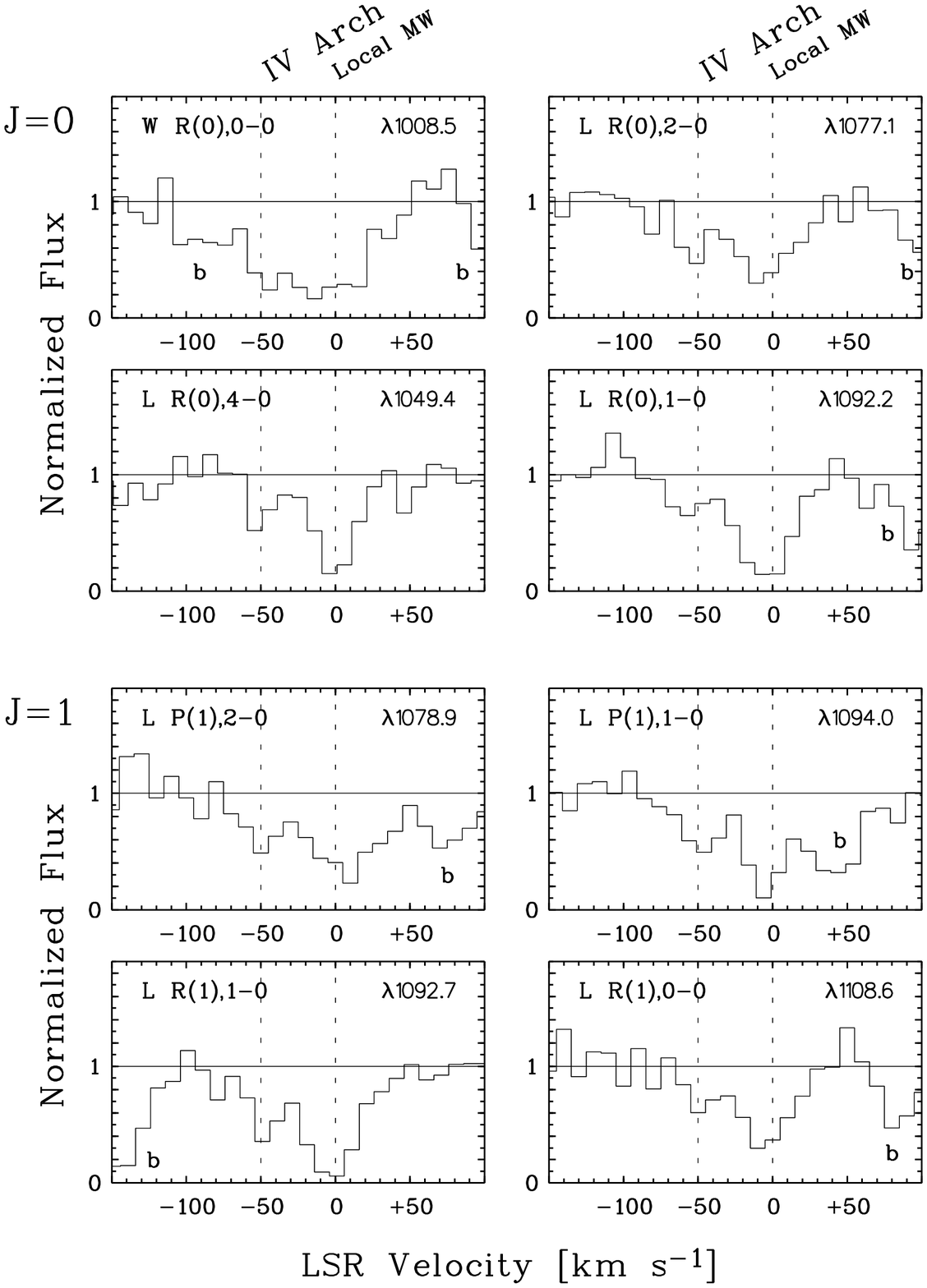}
\figcaption[]{}

\clearpage
\newpage
\includegraphics{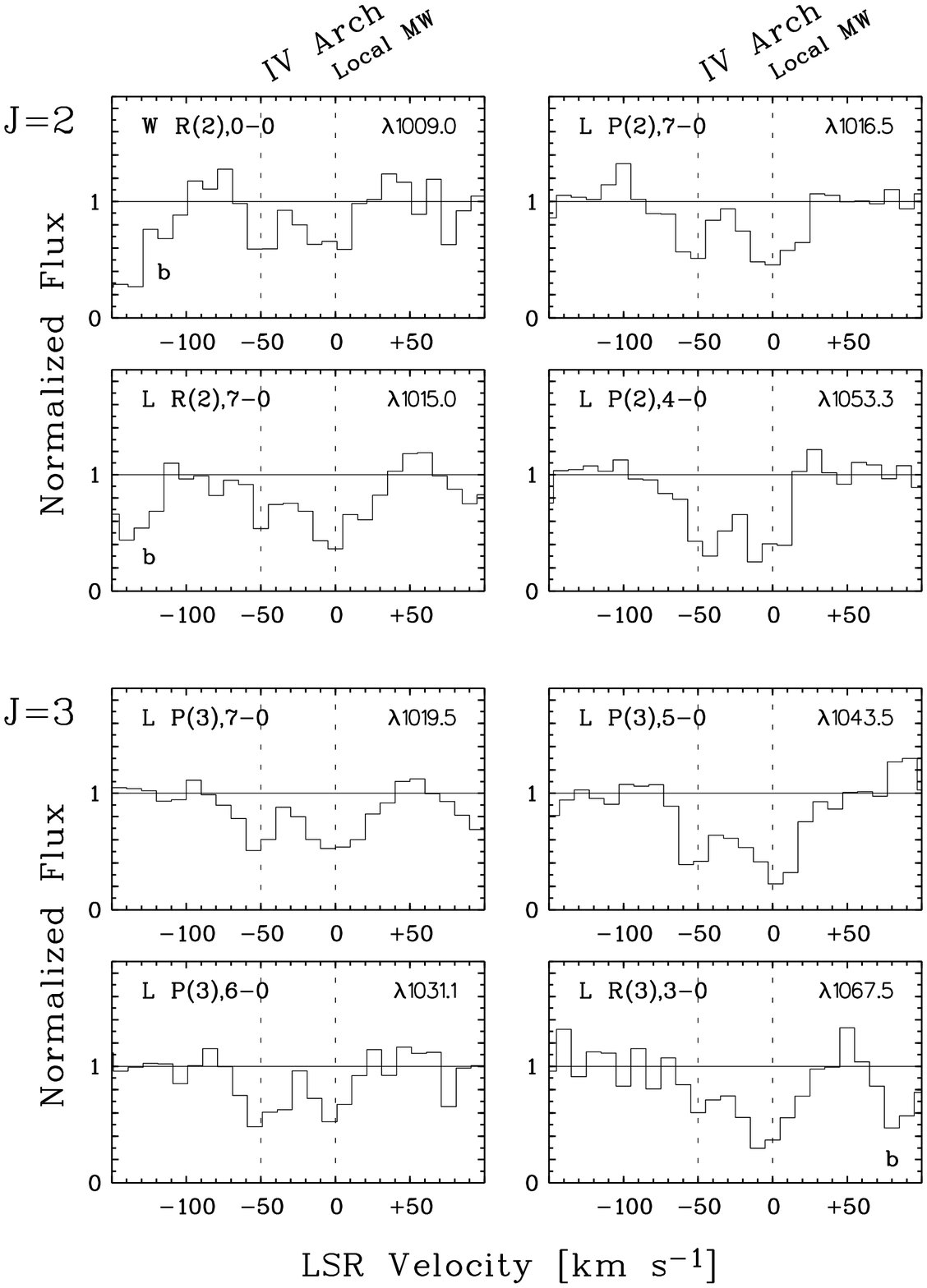}
\figcaption[]{}

\clearpage
\newpage
\includegraphics{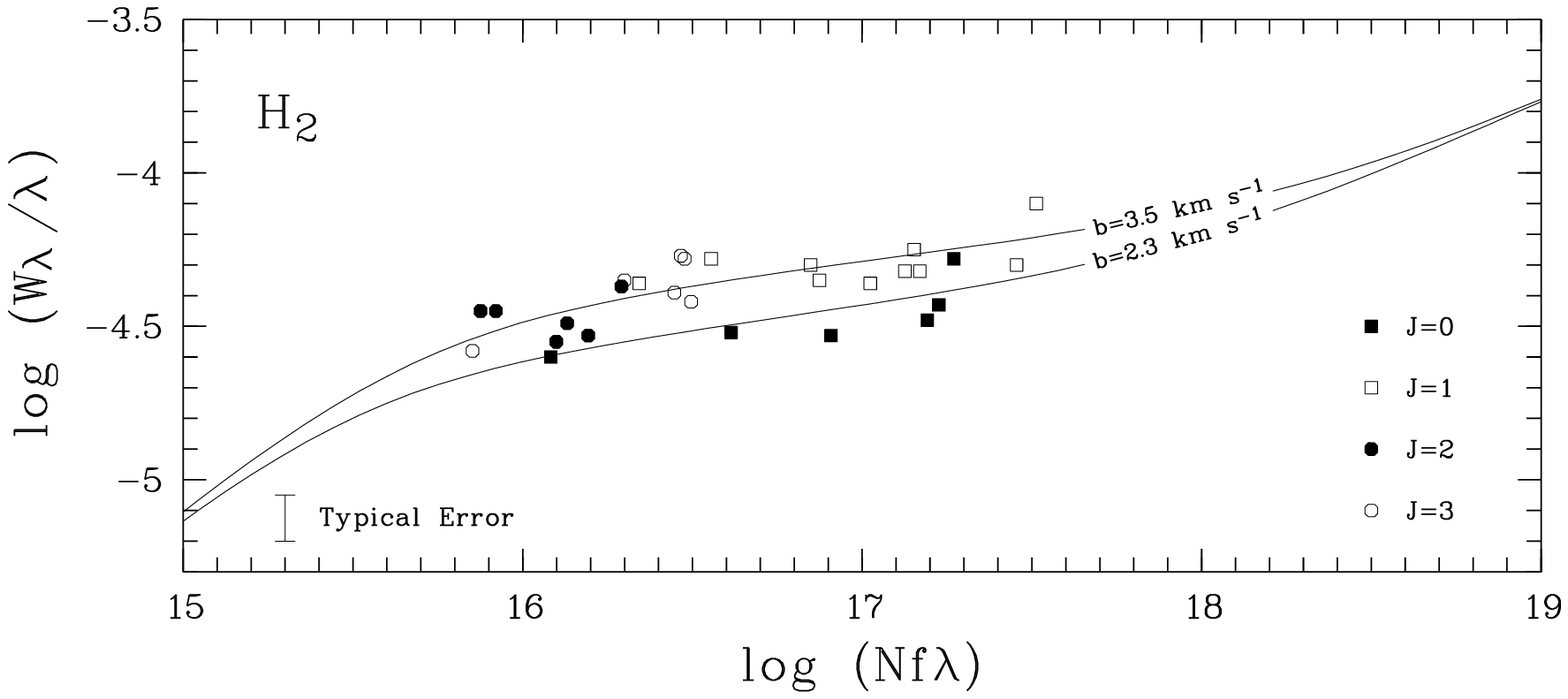}
\figcaption[]{}

\clearpage
\newpage
\includegraphics{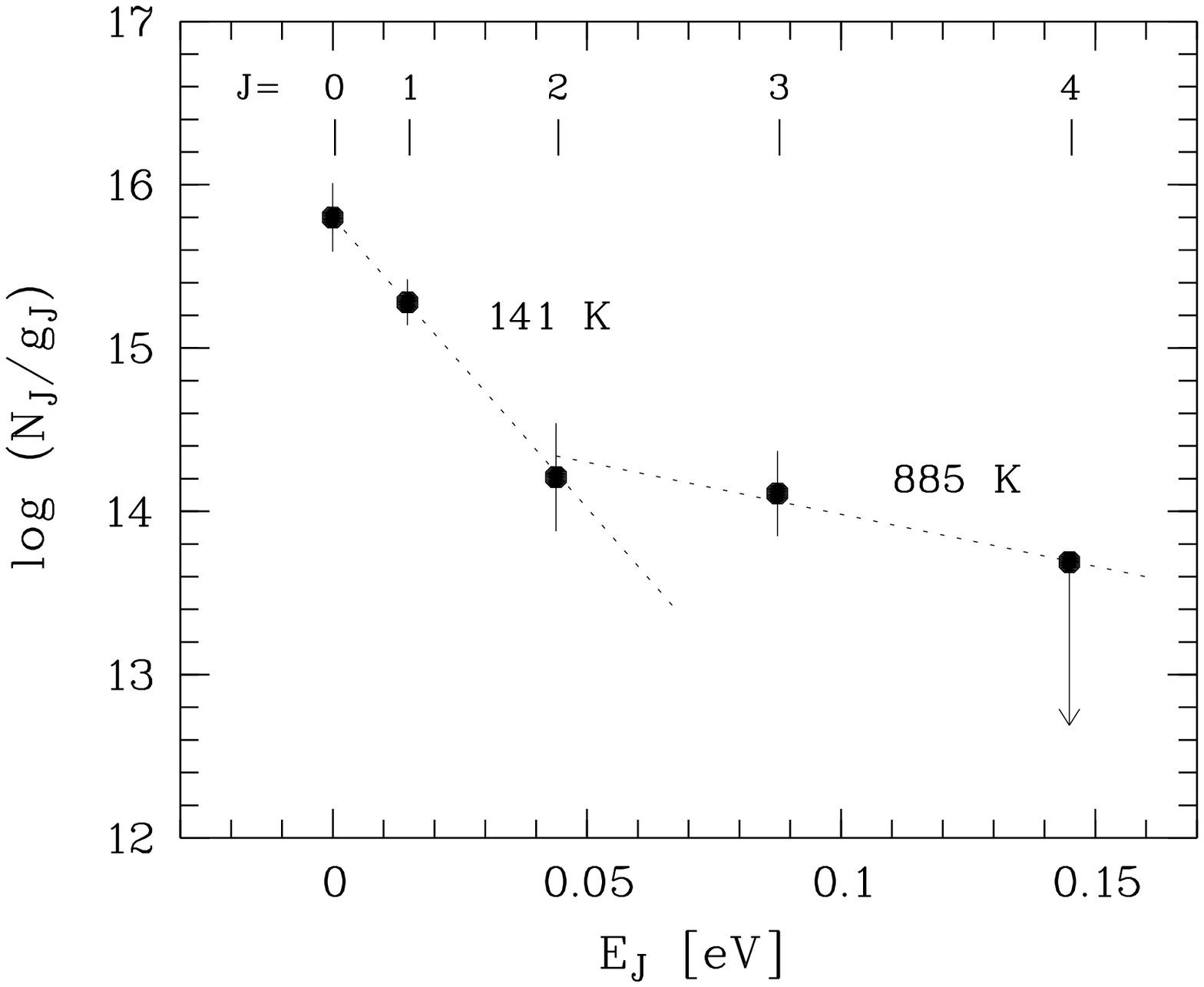}
\figcaption[]{}

\clearpage
\newpage
\includegraphics{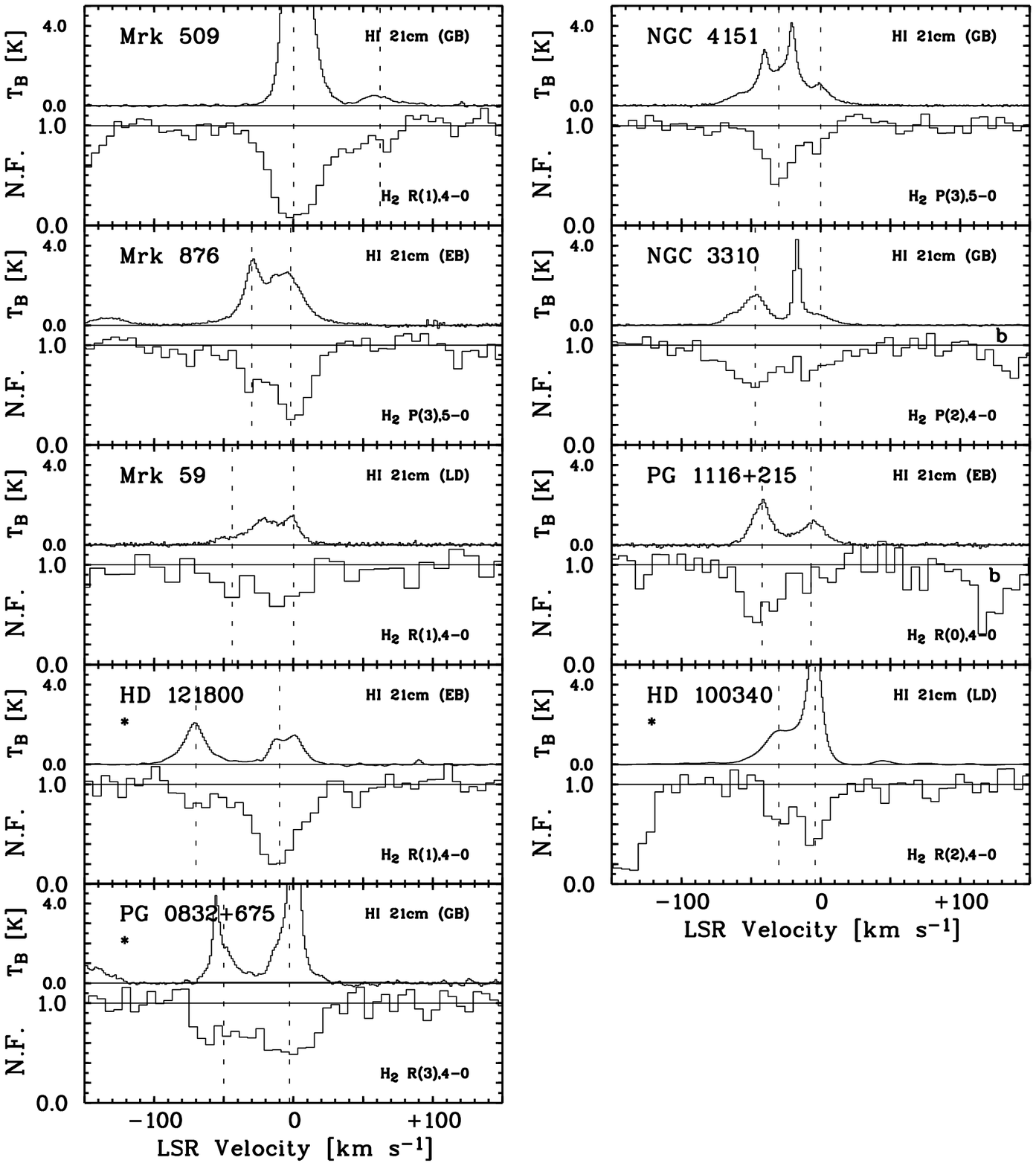}
\figcaption[]{}

\clearpage
\newpage
\includegraphics{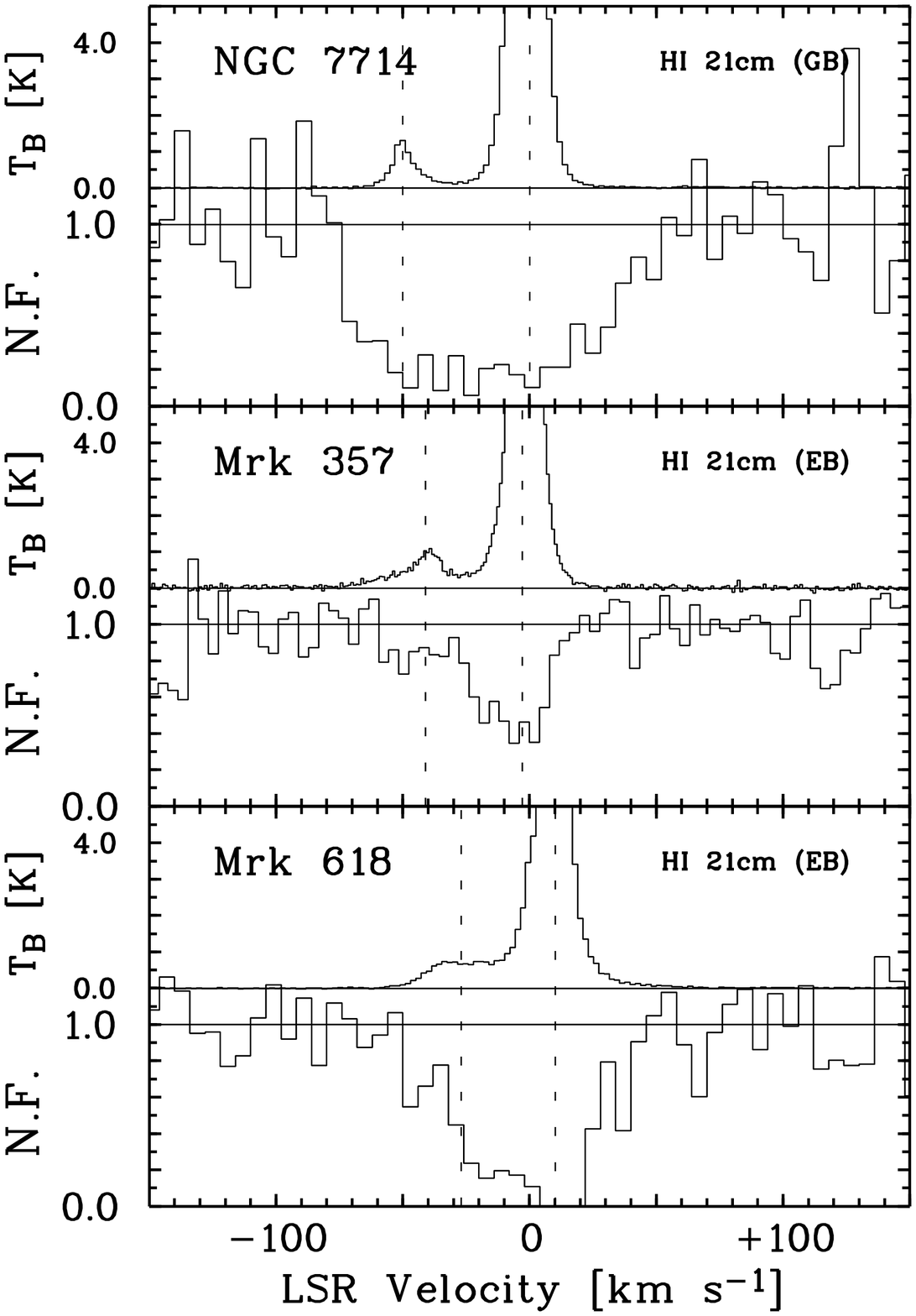}
\figcaption[]{}

\clearpage
\newpage
\includegraphics{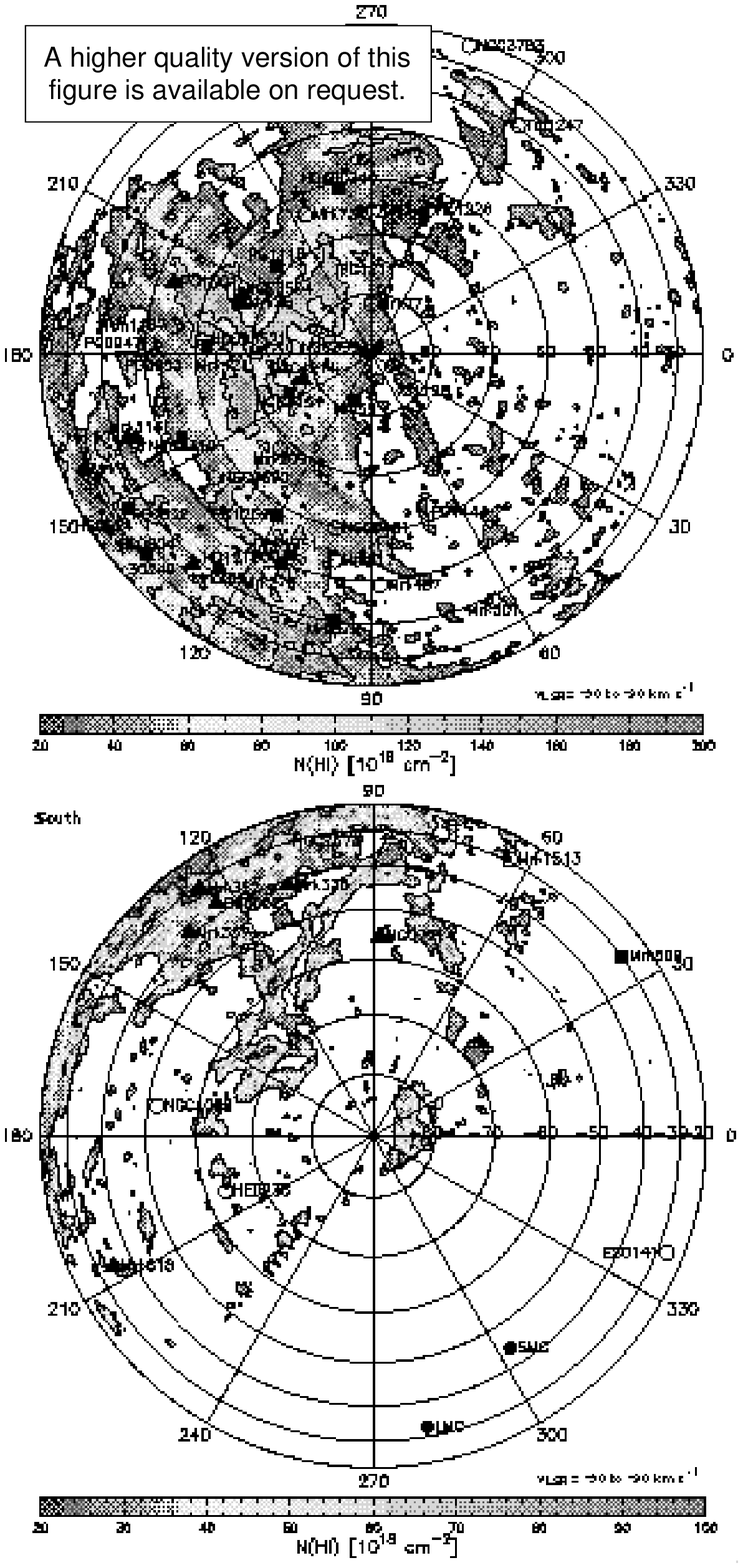}
\figcaption[]{}

\clearpage
\newpage
\includegraphics{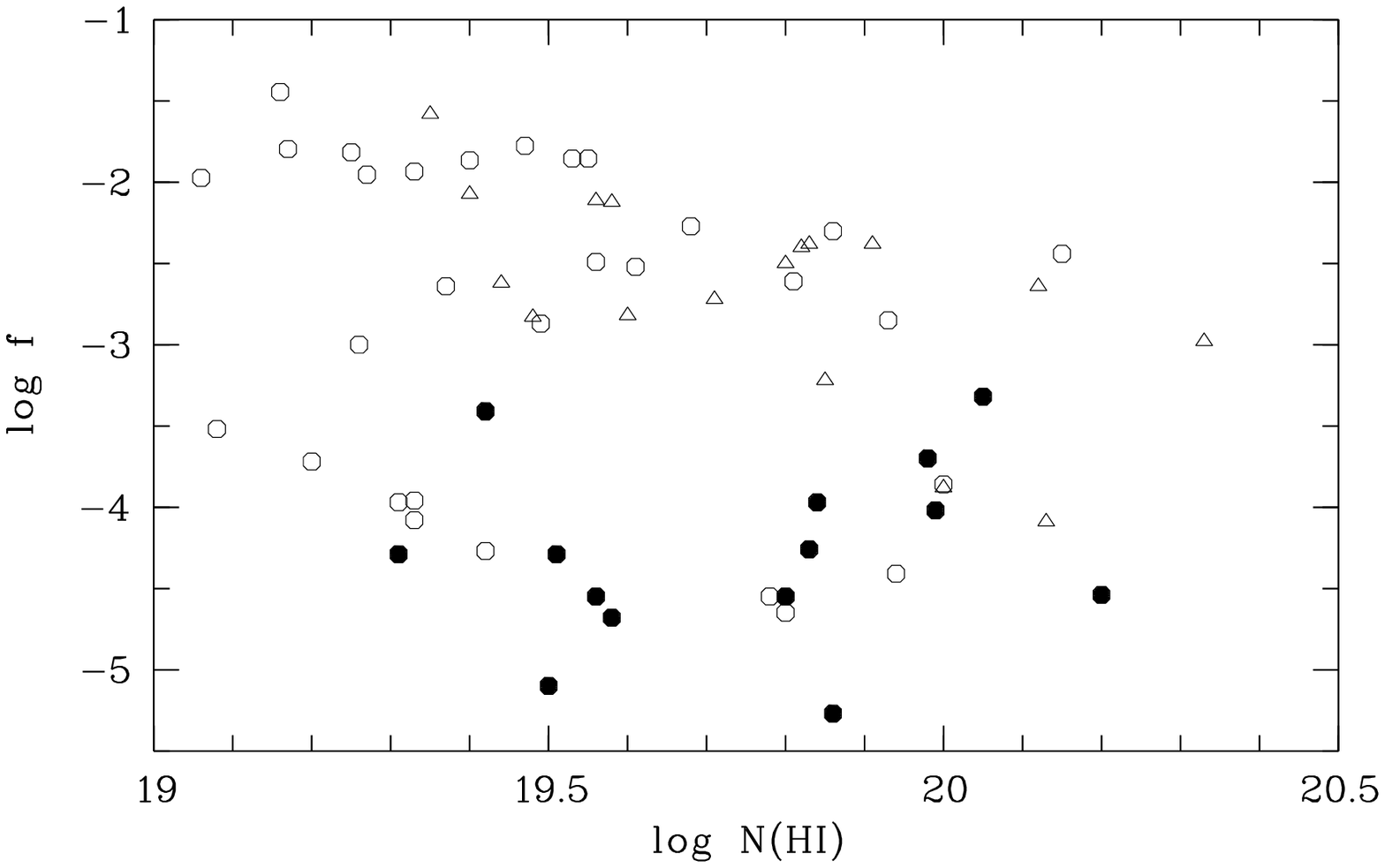}
\figcaption[]{}

\end{document}